%% file: ms.tex
\documentclass{emulateapj}

\usepackage{natbib}
\usepackage{amsmath}
\usepackage{graphicx}
\usepackage[space]{grffile}
\usepackage{latexsym}
\usepackage{amsfonts,amsmath,amssymb}
\usepackage{url}
\usepackage[utf8]{inputenc}
\usepackage{verbatim}
\usepackage{float}
\usepackage{floatflt}
\usepackage{enumitem}

\newcommand{\simgt}{\,\hbox{\lower0.6ex\hbox{$\sim$}\llap{\raise0.6ex\hbox{$>$}}
}\,}
\newcommand{\simlt}{\,\hbox{\lower0.6ex\hbox{$\sim$}\llap{\raise0.6ex\hbox{$<$}}
}\,}

\begin{document}

\title{Mapping the Gas Turbulence in the Coma Cluster: Predictions for {\emph{Astro-H}}}

\author{J. A. ZuHone\altaffilmark{1,2}, M. Markevitch\altaffilmark{2}, I. Zhuravleva\altaffilmark{3,4}}

\altaffiltext{1}{Kavli Institute for Astrophysics and Space Research,
  Massachusetts Institute of Technology, Cambridge, MA 02139}

\altaffiltext{2}{Astrophysics Science Division, X-ray Astrophysics Laboratory,
  Code 662, NASA/Goddard Space Flight Center, Greenbelt, MD 20771}

\altaffiltext{3}{Kavli Institute for Particle Astrophysics and Cosmology,
  Stanford University, 452 Lomita Mall, Stanford, California 94305-4085, USA}

\altaffiltext{4}{Department of Physics, Stanford University, 382 Via Pueblo Mall,
  Stanford, California 94305-4060, USA}

\keywords{galaxies: clusters: intracluster medium --- techniques: spectroscopic --- X-rays: galaxies: clusters --- methods: numerical}

\begin{abstract}
{\it Astro-H} will be able for the first time to map gas velocities and detect turbulence in galaxy clusters. One of the best targets for turbulence studies is the Coma cluster, due to its proximity, absence of a cool core, and lack of a central active galactic nucleus. To determine what constraints {\it Astro-H} will be able to place on the Coma velocity field, we construct simulated maps of the projected gas velocity and compute the second-order structure function, an analog of the velocity power spectrum. We vary the injection scale, dissipation scale, slope, and normalization of the turbulent power spectrum, and apply measurement errors and finite sampling to the velocity field. We find that even with sparse coverage of the cluster, {\it Astro-H} will be able to measure the Mach number and the injection scale of the turbulent power spectrum--the quantities determining the energy flux down the turbulent cascade and the diffusion rate for everything that is advected by the gas (metals, cosmic rays, etc.). {\it Astro-H} will not be sensitive to the dissipation scale or the slope of the power spectrum in its inertial range, unless they are outside physically motivated intervals. We give the expected confidence intervals for the injection scale and the normalization of the power spectrum for a number of possible pointing configurations, combining the structure function and velocity dispersion data. Importantly, we also determine that measurement errors on the line shift will bias the velocity structure function upward, and show how to correct this bias.
\end{abstract}

\bibliographystyle{apj}

\section{Introduction}\label{sec:intro}

X-ray observatories have yielded a wealth of information about the thermodynamic and chemical properties of the intracluster medium (ICM) of galaxy clusters. One aspect of the ICM that has been up to now beyond the ability of present instruments to directly measure is its kinematics. Determining the kinematic properties of the ICM is essential for a complete picture of the physics of the cluster gas. Kinetic energy in the form of bulk motions and turbulence likely provides non-negligible pressure support against gravity, biasing mass estimates based on the assumption of hydrostatic equilibrium, as predicted by simulations \citep{evr96,ras06,nag07,pif08,nel14}, possibly explaining discrepancies between hydrostatic and weak lensing-derived masses \citep{zha10,mah13,vdl14,app14}. Dissipation of turbulent kinetic energy into heat and turbulent transport and mixing of hot gas may partially offset gas cooling in cluster cool cores \citep{fuj04,den05,zuh10,ban14,zhu14}. The properties of gas motions place important constraints on the microphysics of the ICM, in particular its viscosity \citep{fab03,rod13,zuh15}. Finally, ICM turbulence is likely a key ingredient for the origin of non-thermal phenomena such as radio halos and radio mini-halos \citep{bru07,don13,zuh13}.

Though certain observations, such as that of cold fronts (see the review by \citet{MV07} and the recent simulation investigations by \citet{rod13} and \citet{zuh15} for details), indicate that the ICM may be somewhat viscous, the Reynolds number is probably high enough to permit the development of turbulence. Major and minor mergers with other clusters, active galactic nucleus (AGN) outbursts, and stirring of the ICM by the motion of cluster galaxies and dark matter substructure are all possible drivers of turbulent gas motions. In the simplest picture, turbulent motions are driven at a largest, ``injection'' scale, and cascade down to smaller and smaller scales until they are dissipated at the smallest or ``dissipation'' scale.

In this picture, the distinction between ``bulk flows'' and ``turbulence'' is blurry--most merger-induced large-scale bulk flows simply represent turbulence at its injection scale. A ``bulk flow'' in a merging cluster could not be considered part of the turbulent field if the moving gas parcel is still gravitationally bound to an infalling subcluster (i.e., it is still part of the ``spoon'' that stirs the turbulence);  when that gas is stripped by ram pressure, it joins the turbulent velocity field. Infalling subclusters are usually easy to identify in the X-ray images and, if necessary, can be masked for a study of the turbulent velocities.

Physical models of the plasma combined with dimensional arguments yield simple power-law forms for the turbulent power as a function of wavenumber, the most familiar of which is the Kolmogorov spectrum. Therefore, observationally determining the relationship between measured gas velocities and the range of length scales in the cluster has the potential to reveal the underlying gas physics. In particular, determining the dissipation scale of the turbulent cascade would give the effective ``viscosity'' of the ICM, which is governed by complex dissipation processes in the magnetized, collisionless plasma. Knowledge of the injection scale of the cascade, in combination with the average velocity dispersion (or Mach number) of the turbulent motions, would give us the diffusion rate of everything that is advected by the gas \citep[such as metals, cosmic rays, specific entropy,][]{reb06,vaz10,ens11}, because the largest-scale eddies are those most responsible for stirring and mixing the ICM. This combination also determines the flux of energy that moves down the turbulent cascade and ends up dissipating into heat and accelerating cosmic rays \citep{den05,bru07}.

Turbulence in the ICM can be studied directly by several methods. \citet{ino03} pointed out that the finite number of the largest-scale turbulent eddies that fit in a cluster volume will result in deviations of the spectral line profiles from a simple Gaussian shape. This method requires well-resolved spectra with high statistical quality to detect subtle shape deviations. Another method relies on mapping the projected line of sight velocities over the face of a cluster to derive a velocity structure function, which has a one-to-one correspondence to the turbulent power spectrum \citep[][hereafter Z12, and references therein]{zhu12}. This method is less demanding to the statistical accuracy of the line profiles--only the line positions are used. We will consider this method below.

So far, X-ray observatories have lacked the required spectral resolution to measure the line shifts and broadening resulting from turbulent motions in the ICM. The RGS grating on {\it XMM-Newton} can provide weak upper limits on Doppler broadening of spectral lines in cool-core clusters \citep[][and references therein]{san11,san13a,pin15}. Indirect estimates of the ICM turbulent velocity can be obtained from measurements of resonant scattering \citep[e.g.,][]{chu04,wer09,dep12,zhu13}, pressure fluctuations \citep{sch04}, or surface brightness fluctuations \citep{chu12,zhu15}.

The {\it Astro-H} Mission, a joint JAXA/NASA endeavor, will be launched in late 2015 and will for the first time have the energy resolution to detect line shifts and widths resulting from bulk and turbulent motions in the ICM \citep{tak12}. {\it Astro-H} will possess a Soft X-ray Spectrometer (SXS) micro-calorimeter with an energy resolution of $\Delta{E} \leq 7$~eV within the energy range $E \sim 0.3-12.0$~keV, covering a 3'$\times$3' field. At the energy of the Fe-K$_{\alpha}$ line, $E \approx 6.7$~keV, this enables the measurement of velocities at resolutions of tens of km~s$^{-1}$.

\begin{figure*}
\begin{center}
\includegraphics[width=0.95\textwidth]{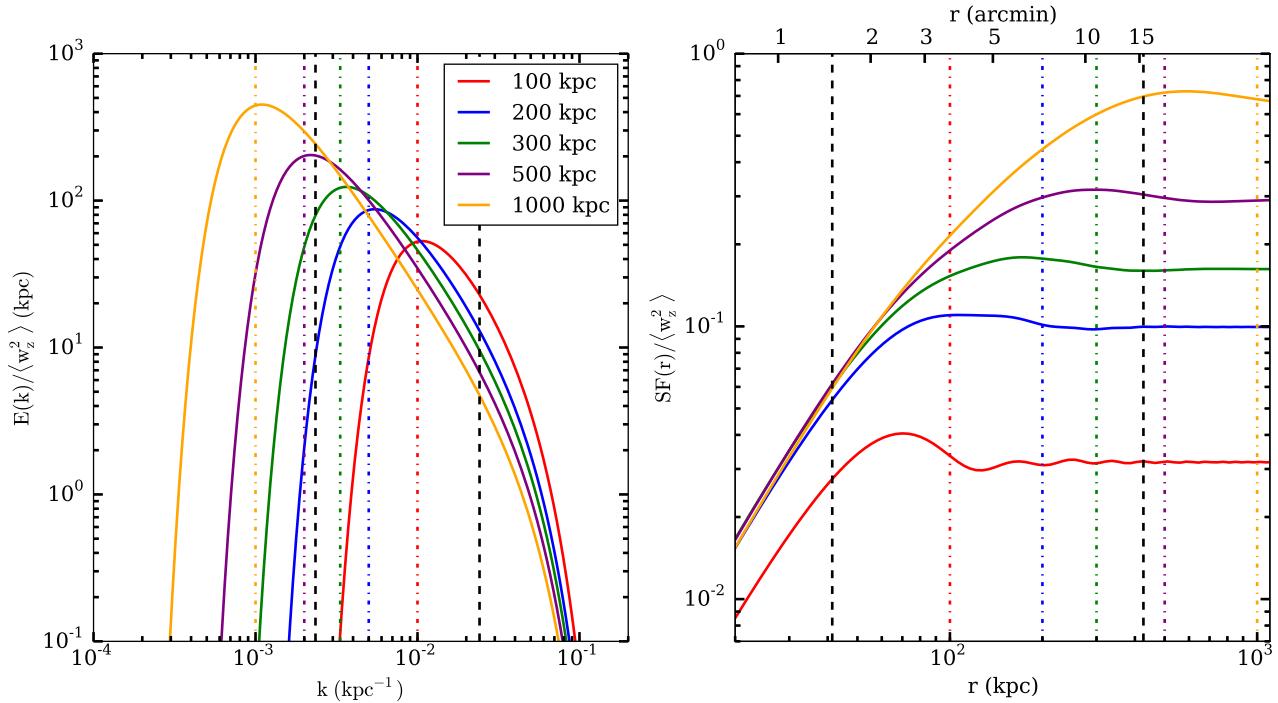}
\caption{Theoretical energy spectra and corresponding structure functions for a Coma-like cluster, calculated without measurement or ``cosmic variance'' uncertainty for different large-scale cutoffs (injection scales) of the velocity power spectrum. The dissipation scale for all curves is 20~kpc, and all spectra have $\alpha$ = -11/3. Colored dot-dashed lines indicate the values of the cutoff scale for each of the different energy spectra. The curves have been normalized by the average velocity dispersion $\langle{w_z^2}\rangle$, which is the integral of $P_{\rm 3D}$ over all $k$. Dashed black lines indicate the lowest and highest scales resolvable by the pointing configurations in Figure \ref{fig:pattern_shapes}.}
\label{fig:semi_analytic_upper}
\end{center}
\end{figure*}

A promising target for measuring the kinematic properties of the ICM with {\it Astro-H} is the Coma cluster. Coma is nearby ($\sim$~100 Mpc, $z$ = 0.0231), possesses a flat core with a radius of $r_c \sim 300$~kpc ($\approx$11'), and is likely to be turbulent due to ongoing merger activity. Further, Coma lacks a central AGN, which often generate ICM motions in their vicinity. Its lack of strong entropy gradients, unlike those in cool-core clusters, implies that the effects of stratification are not significant, and turbulence should develop in an isotropic, qualitatively simple way. We do note that \citet{san13b} argued that the presence of large-scale filamentary structures they have detected in the X-ray image of Coma implies that turbulence is suppressed. However, such large-scale coherent structures are seen in hydrodynamic simulations in turbulent clusters \citep[e.g.][]{vaz09}, so a comparison with simulations is needed to draw conclusions from those observations.

In this work, we will simulate turbulent velocity fields in the Coma cluster, using a simple form for the underlying power spectrum to test the ability of {\it Astro-H} to constrain the slope of the power spectrum and relevant length scales, such as the injection scale of the turbulent motions and the dissipation scale. From these simulated velocity fields, we will construct the second-order structure function, a measure of the spatial structure of the velocity field that is equivalent to the power spectrum, and fit it with various models, in combination with the velocity dispersion. We will investigate a number of possible spatial configurations for individual SXS pointings, in order to determine which are most useful for constraining particular aspects of the velocity field. We assume a flat $\Lambda$CDM cosmology with $h$ = 0.71 and $\Omega_{\rm m}$ = 0.27. At the distance of Coma, an angular size of 1' corresponds to a length scale of $\sim$26~kpc.

\section{Method}\label{sec:method}

In what follows, we closely follow the work of Z12, who derived a number of important formulae for determining the properties of gas motions in the ICM from measurements of the shift and width of spectral lines. Our particular case of the Coma cluster allows us to make some further simplifying assumptions, as we will note below.

\subsection{Cluster Model}\label{sec:model}

We construct a simple model of the Coma cluster assuming that the electron number density follows a $\beta$-model \citep{cav76,cav78}:
\begin{equation}\label{eqn:beta_model}
n_e(r) = n_0\left[1+\left(\frac{r}{r_c}\right)\right]^{-3\beta/2}
\end{equation}
with $n_0$ = 0.003~cm$^{-3}$, $r_c$ = 300~kpc, and $\beta$ = 2/3. We assume that the gas is isothermal with $T$ = 8~keV and has constant metallicity $Z = 0.3~Z_\odot$. These parameters are chosen to provide a simple model with which to perform our calculations, and are broadly consistent with $\beta$-model fits to the Coma cluster emission \citep[e.g.,][]{hug89,bri92,neu03}

Without any loss of generality, throughout this work we will assume the line of sight is along the $z$ direction, the 3D position vector ${\bf x} = (x,y,z)$, and the 3D wavevector ${\bf k} = (k_x,k_y,k_z)$.\footnote{Here and throughout we adopt the convention that wavenumbers $k$ and distance scales $r$ are related by $k = 1/r$.} The 3D velocity field $v_z(\bf{x})$ is assumed to be turbulent with an underlying power spectrum
\begin{equation}\label{eqn:power_spec}
P_{\rm 3D}(k) = |\tilde{v}_z({\bf k})|^2 = C_n{e^{-(k_1/k)^2}k^{\alpha}e^{-(k/k_0)^2}}
\end{equation}
where $\tilde{v}_z$ is the Fourier transform of $v_z$, $k = |{\bf k}|$, $k_0$ and $k_1$ are cutoffs of the power spectrum at high and low wavenumber, respectively, $C_n$ is a normalization constant, and $\alpha$ is the spectral index. The corresponding energy spectrum is given by:
\begin{equation}
E(k) \sim P_{\rm 3D}(k){k^2}
\end{equation}
For a Kolmogorov power spectrum, $\alpha = -11/3$ results in the
familiar $k^{-5/3}$ behavior for the energy spectrum in the
inertial range of Equation \ref{eqn:power_spec}.

\begin{figure*}
\begin{center}
\includegraphics[width=0.95\textwidth]{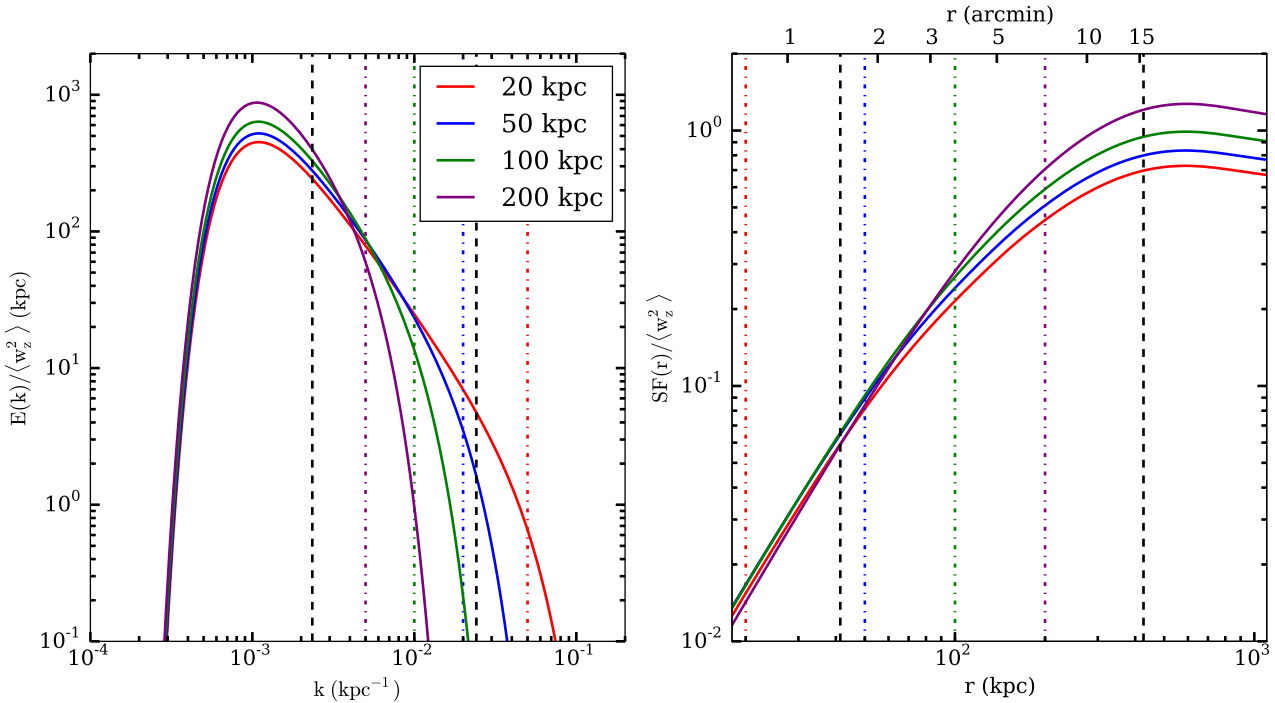}
\caption{Theoretical energy spectra and corresponding structure functions for a Coma-like cluster, calculated without measurement or ``cosmic variance'' uncertainty for different small-scale cutoffs (dissipation scales) of the velocity power spectrum. The injection scale for all curves is 1000~kpc, and all spectra have $\alpha$ = -11/3. Colored dot-dashed lines indicate the values of the cutoff scale for each of the different energy spectra. The curves have been normalized by the average velocity dispersion $\langle{w_z^2}\rangle$, which is the integral of $P_{\rm 3D}$ over all $k$. Dashed black lines indicate the lowest and highest scales resolvable by the pointing configurations in Figure \ref{fig:pattern_shapes}.}
\label{fig:semi_analytic_lower}
\end{center}
\end{figure*}

\subsection{Computing Line of Sight Velocities}\label{sec:los_vels}

The emission-weighted line of sight velocity field is given by:
\begin{equation}\label{eqn:los_vel1}
\bar{v}_z(x,y) = \frac{\int{v_z({\bf x})}\epsilon({\bf x})dz}{\int{\epsilon({\bf x})dz}}
\end{equation}
where $\epsilon$ is the X-ray volume emissivity $n_en_H\Lambda(T,Z)$, with $n_e$, $n_H$ the electron and proton number densities, $T$ the plasma temperature, and $Z$ the metallicity. The assumptions of isothermality and constant metallicity render $\bar{v}_z$ dependent on $n_en_H$ alone. To simplify the equations, we set (as in Z12):
\begin{equation}\label{eqn:EM}
\omega({\bf x}) = \frac{\epsilon({\bf x})}{\int{\epsilon({\bf x})dz}}
\end{equation}
resulting in
\begin{equation}\label{eqn:los_vel2}
\bar{v}_z(x,y) = \int{v_z({\bf x})}{\omega({\bf x})dz}
\end{equation}
The line of sight velocity dispersion is similarly computed by:
\begin{equation}\label{eqn:los_disp}
w_z^2(x,y) = \int{v_z^2({\bf x})}{\omega({\bf x})dz} - \bar{v}_z^2(x,y)
\end{equation}

\subsection{2D Power Spectrum and Structure Function}\label{sec:2d_ps_and_sf}

Following Z12 (see their Section 3, Equation 4; also Appendices A and B) and \citet{chu12} (see their Section 3), a relationship between the power spectrum of the 3D velocity field and that of the 2D velocity field can be derived:
\begin{equation}\label{eqn:3d_to_2d}
P_{\rm 2D}(k_x,k_y;x,y) = \displaystyle\int{P_{\rm 3D}}({\bf k})P_\epsilon(k_z;x,y)dk_z
\end{equation}
where $P_{\rm 2D}$ is the power spectrum of the line of sight velocity field $\bar{v}_z$ and $P_\epsilon$ is the power spectrum of the normalized emission measure along the z-axis. They also pointed out that for a large wavenumbers $k$, the relationship between the 2D and 3D velocity power spectra is approximately independent of the integral over the power spectrum of the emission measure (cf. Section 6 of Z12, in particular Equations 17-19 and Figure 8):
\begin{equation}
P_{\rm 2D}(k) \approx P_{\rm 3D}(k)\displaystyle\int{P_\epsilon(k_z;x,y)}dk_z = KP_{\rm 3D}(k)
\end{equation}
e.g., the 2D power spectrum is the same as the 3D power spectrum apart from a normalization constant, which is easily computed for our model cluster (see Appendix \ref{sec:norm} for a derivation).

In our analysis, we will use the simulated velocity fields to compute the second-order structure function:
\begin{equation}\label{eqn:2d_sf}
{\rm SF}(r) = \langle|\bar{v}_z(\boldsymbol{\chi}+{\bf r})-\bar{v}_z(\boldsymbol{\chi})|^2\rangle,
\end{equation}
where $\boldsymbol{\chi}$ and $\bf{r}$ are two-dimensional position vectors on the sky, $r = |\bf{r}|$, and $\langle\rangle$ indicates an average over pairs of points with the same separation $r$. This quantity has the advantage that it can be computed easily regardless of the spatial shape of the region under consideration. The second-order structure function is directly related to the power spectrum of the projected velocity field (see Appendix \ref{sec:sf_derivation} for a derivation):
\begin{equation}\label{eqn:ps_to_sf}
SF(r) = 4\pi\int_0^\infty{\left[1-J_0(2\pi{kr})\right]P_{\rm 2D}(k)kdk}
\end{equation}
where $J_0$ is a Bessel function of the first kind.

If we only consider the power-law form of the power spectrum, we can derive a scaling for the structure function in the inertial range. Neglecting the variation in $J_0$, and assuming a characteristic scale of $k_c = 1/r$ for the turbulent cascade, we have
\begin{equation}\label{eqn:alpha_to_gamma}
SF(r) \propto \int_{k_c}^\infty{P_{\rm 2D}kdk} \propto k_c^{\alpha+2} \propto r^\gamma,
\end{equation}
with $\gamma = -(\alpha+2)$. For a Kolmogorov spectrum with $\alpha = -11/3$, $\gamma = 5/3$.

To determine what form we may expect $SF(r)$ to take for different forms of the power spectrum $P_{\rm 2D}(k)$, we will numerically integrate Equation \ref{eqn:ps_to_sf}, assuming a Coma-like cluster and ignoring sources of error. We choose to examine two sets of models, one with different cutoffs of the power spectrum at small scales ($k_0 = 1/l_{\rm min}$), and another with different cutoffs of the power spectrum at large scales ($k_1 = 1/l_{\rm max}$). The former represents different possible values of the dissipation scale of the turbulent motions, whereas the latter represents different possible values of the injection scale of the turbulent motions. Each set will have the same cutoff scale at the opposite end. Figures \ref{fig:semi_analytic_upper} and \ref{fig:semi_analytic_lower} show the energy spectra and structure functions calculated for the large-scale and small-scale cutoffs, respectively, assuming a Kolmogorov slope of $\alpha = -11/3$. The black dashed lines in the figure indicate the range of length scales sampled by the observations we will simulate in Section \ref{sec:simulations}. The structure functions with the small-scale cutoffs (left panel) all have a very similar shape over the relevant length scales, diverging only slightly toward the smaller scales. On the other hand, the curves with the large-scale cutoffs exhibit large differences in shape over the relevant length scales. As the cutoff scale $l_{\rm max}$ is decreased, the structure function becomes flatter, and with $l_{\rm max} = 100$~kpc the curve is essentially constant over all sampled length scales. The structure functions are normalized by the average velocity dispersion (which will be measured independently from the line width), in order to illustrate the extra information contained in the shape of the structure function itself.

\subsection{Defining Parameter Spaces}\label{sec:defining_params}

To decide on our chosen values of the various input parameters for our velocity power spectra, we draw on both theoretical and observational constraints specific to the Coma cluster.

We determine our range for the injection scale based on observations of the Coma cluster and theoretical considerations. \citet{chu12} used {\it Chandra} and {\it XMM-Newton} observations to analyze surface brightness fluctuations in Coma, deriving a 3D power spectrum of density fluctuations. On large scales, the spectrum of fluctuations begins to turn over for length scales larger than $l_{\rm max} \sim 300-500$~kpc (their Figures 14 and 15). If the density fluctuations are associated with turbulence, we can identify this scale with the injection scale of the gas motion. Coma's large radio halo \citep{bro11} and lack of a central cool core are evidence for recent merging activity, which can drive gas motions on scales up to a Mpc \citep{ric01,poo06}. A recent simulation of the formation of a Coma-like cluster by \citet{min15} estimated an injection scale of $\sim$1~Mpc. Galaxies moving through the ICM can drive turbulence at smaller scales, around $\sim$100~kpc. From these considerations, we adopt a wide range of injection scales from 100~kpc up to 1000~kpc for our simulations.

For the dissipation scale, we may use the \citet{kol41} formalism to estimate what a reasonable expectation of its value may be. Assuming turbulence is injected at scale $l_{\rm max}$ with velocity $v_{\rm max}$, we may estimate the energy injection rate as
\begin{equation}\label{eqn:inj_rate}
\epsilon \sim v_{\rm max}^3/l_{\rm max}
\end{equation}
this energy will be dissipated at the scale $l_{\rm min}$ with a characteristic velocity $v_{\rm min}$ given by
\begin{equation}\label{eqn:Re}
l_{\rm min}v_{\rm min} \sim \nu
\end{equation}
where $\nu$ is the kinematic coefficient of viscosity. Since in the inertial range the energy cascade rate is a constant, we may combine Equations \ref{eqn:inj_rate} and \ref{eqn:Re} to give the dissipation scale as
\begin{equation}
l_{\rm min} = \left(\frac{\nu^3}{\epsilon}\right)^\frac{1}{4}
\end{equation}
If we assume the viscosity is given by the Spitzer \citep{spi62} formulation, given the properties of the plasma given in Section \ref{sec:model} and the range of velocities and injection scales that we assume, we derive a range for the dissipation scale of the turbulence of $\l_{\rm min} \sim 10-50$~kpc. The entire range for the dissipation scale is smaller than or comparable to the {\it Astro-H} resolution at the redshift of Coma, implying that it will be difficult to constrain this parameter. For this reason, we have not only chosen values for the dissipation scale $l_{\rm min}$ within this range, but also somewhat higher (100-200~kpc), to determine if {\it Astro-H} would be able to distinguish even extreme values of this scale from more physically sensible ones.

\begin{figure*}
\centering
\includegraphics[width=0.3\textwidth]{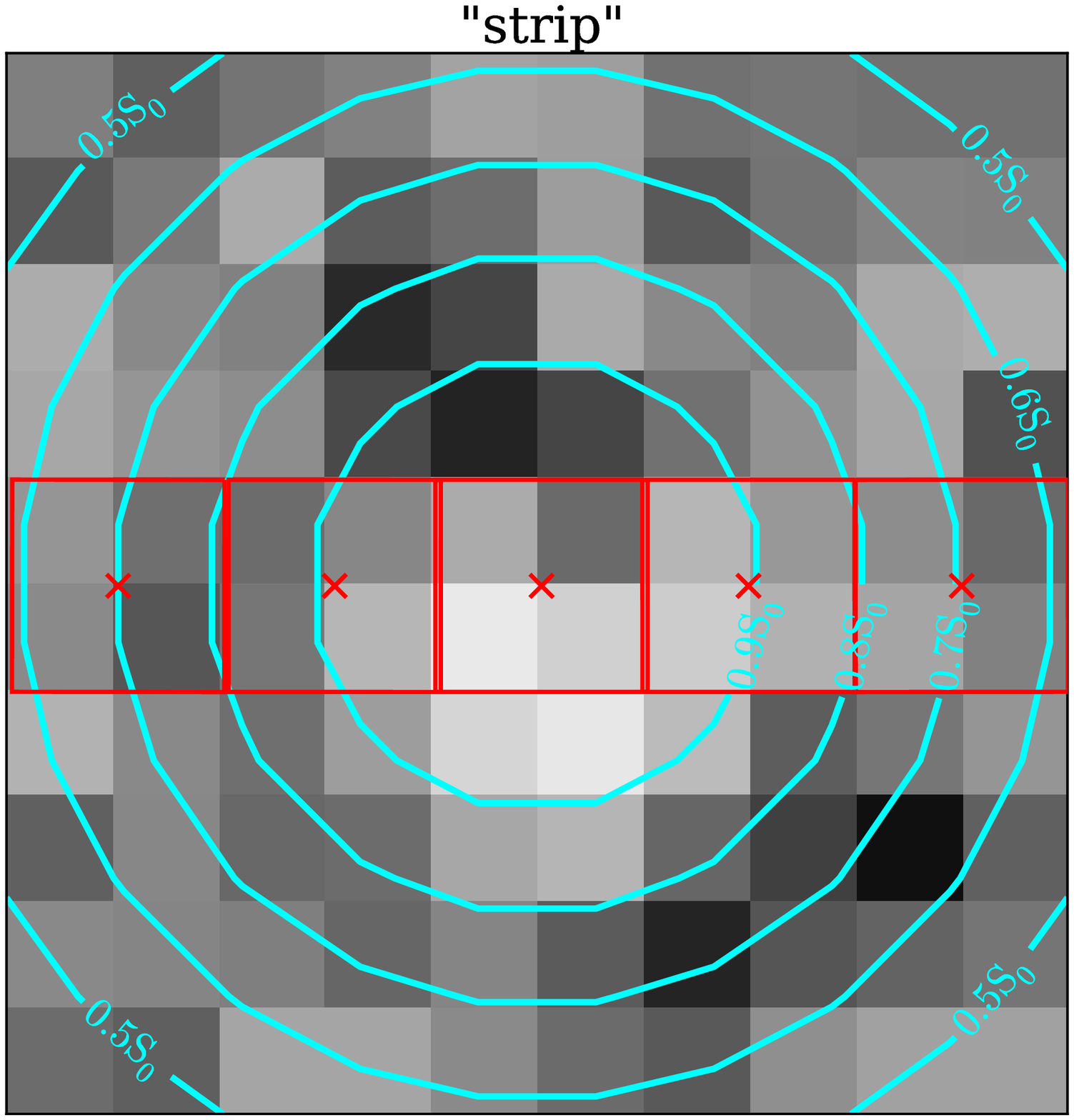}
\includegraphics[width=0.3\textwidth]{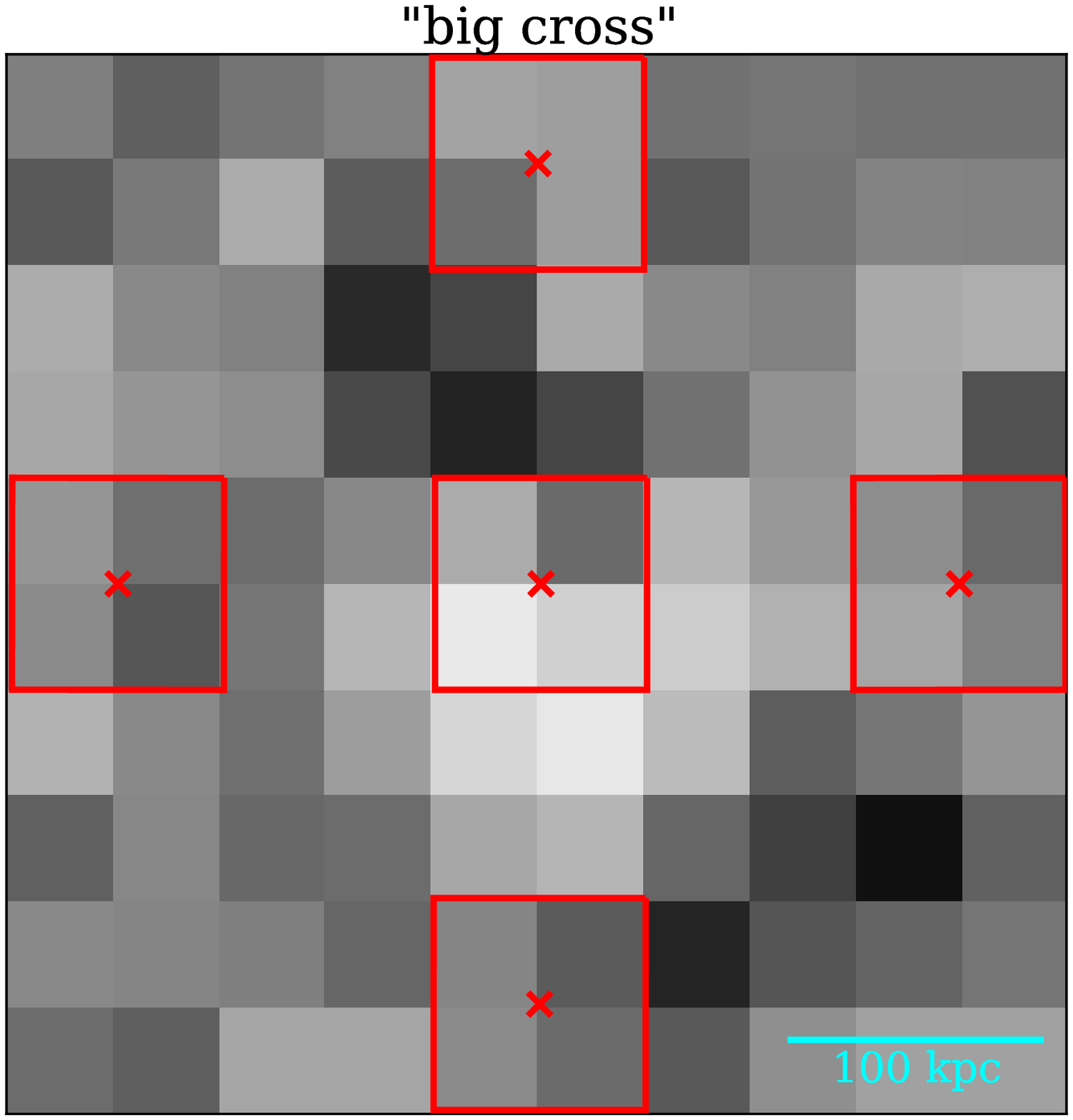}
\includegraphics[width=0.3\textwidth]{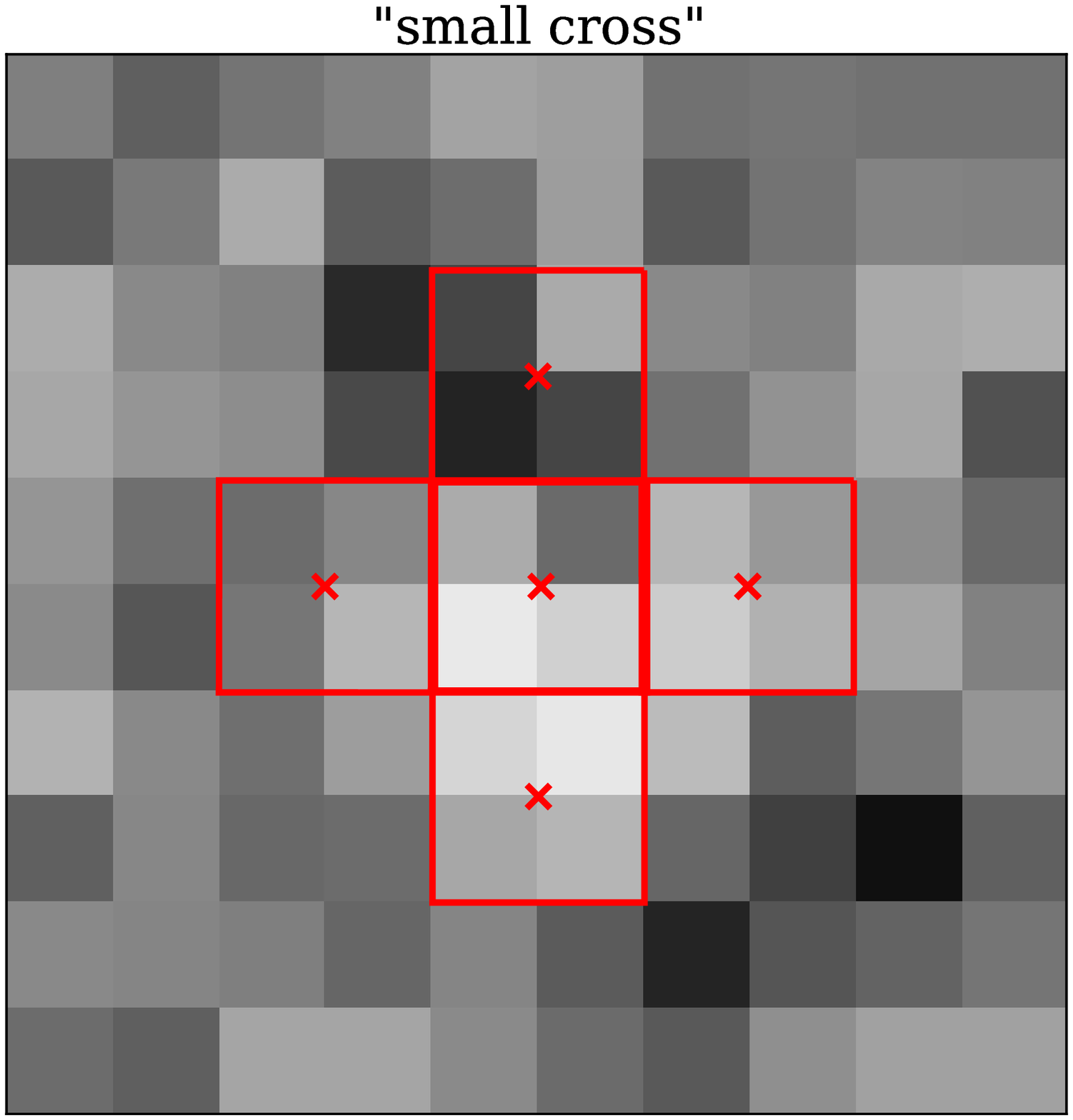}
\includegraphics[width=0.3\textwidth]{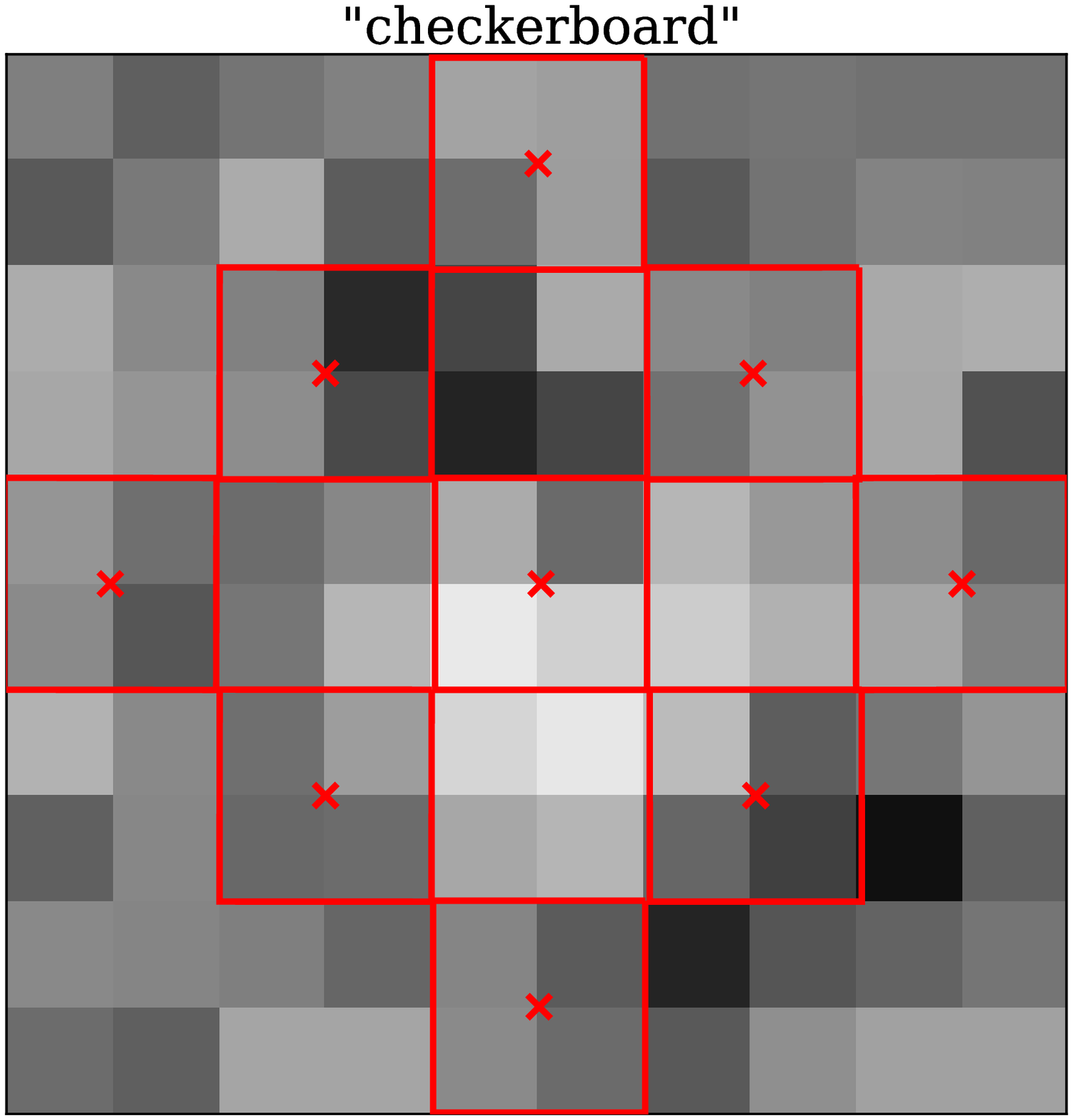}
\includegraphics[width=0.3\textwidth]{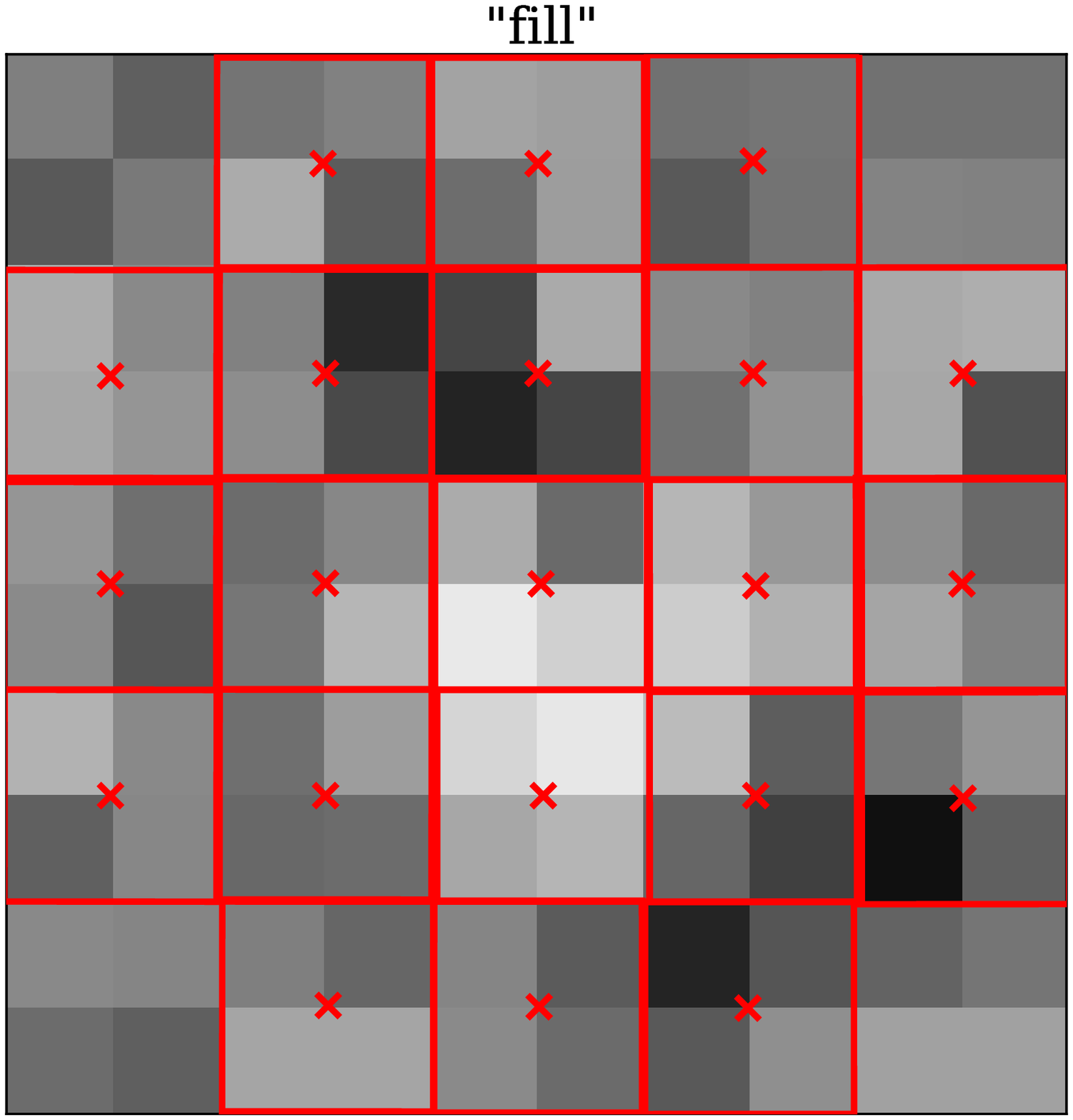}
\caption{Configurations over which the structure function is calculated, shown as combinations of red 3'$\times$3' squares, overlaid on a sample line centroid field. Contours in the upper-left panel show the X-ray surface brightness distribution in units of $S_0$, the central surface brightness.}
\label{fig:pattern_shapes}
\end{figure*}

The choice of an isotropic Spitzer viscosity constitutes a likely upper limit on the viscosity of the ICM. Given that the cluster plasma is weakly magnetized, the ion mean free path is much larger than its Larmor radius, which renders the viscosity anisotropic \citep{bra65}. Under the assumption of an isotropically tangled magnetic field, this will suppress the average viscosity by a factor of 5 \citep{nul13}. A proper treatment of turbulent dissipation in such a magnetized medium would take into account the fact that only fast and slow magnetosonic modes will be damped by Braginskii viscosity, whereas Alfvenic modes will be undamped \citep{bru07}. The effective viscosity of the ICM on small scales may depend on whether or not the turbulence is super or sub-Alfvenic \citep{laz06a,laz06b}. Finally, microscale plasma instabilities (such as firehose and mirror) may limit viscous damping caused by ion collisions, reducing the viscosity and the dissipation scale even further \citep{laz06,sch06,kun14,mog14}. These complex issues are beyond the scope of this paper, and beyond the capabilities of {\it Astro-H} to explore, so we use the Spitzer viscosity as a useful benchmark.

Finally, in choosing the spectral index $\alpha$ of the turbulent cascade, we rely on different theoretical descriptions of turbulence that are relevant for galaxy clusters. If the turbulence is primarily hydrodynamic and incompressible, we expect the standard Kolmogorov scaling with $\gamma = 5/3$ (as defined in Equation \ref{eqn:alpha_to_gamma}). If the turbulence is compressible with weak shocks in an otherwise subsonic velocity structure, we expect a Burgers scaling with $\gamma$ = 2. Alternatively, if the turbulence is magentohydrodynamical in nature, we expect a Kraichnan spectrum with $\gamma$ = 3/2 \citep[see][for a review]{bra13}.

\subsection{Generating Simulated Velocity Fields and their Structure Functions}\label{sec:simulations}

Using the form of the 3D power spectrum defined by Equation \ref{eqn:power_spec}, we can generate realizations of the 3D velocity field. Each realization of the velocity field is set up on a uniformly gridded rectangular domain with a projected area on the sky of size ${L_x}\times{L_y} = 15'\times15'$, or $\approx 414$~kpc on a side, and a line of sight width $L_z = 2070$~kpc. The domain is subdivided into with ${n_x}\times{n_y}\times{n_z} = 40\times40\times200$ cells, resulting in a cell width of $\approx 10$~kpc on a side. This resolution is sufficient for our purposes, as it is several times smaller than the {\it Astro-H} resolution at the distance of Coma, approximately 40~kpc.

\begin{figure*}
\begin{center}
\includegraphics[width=0.30\textwidth]{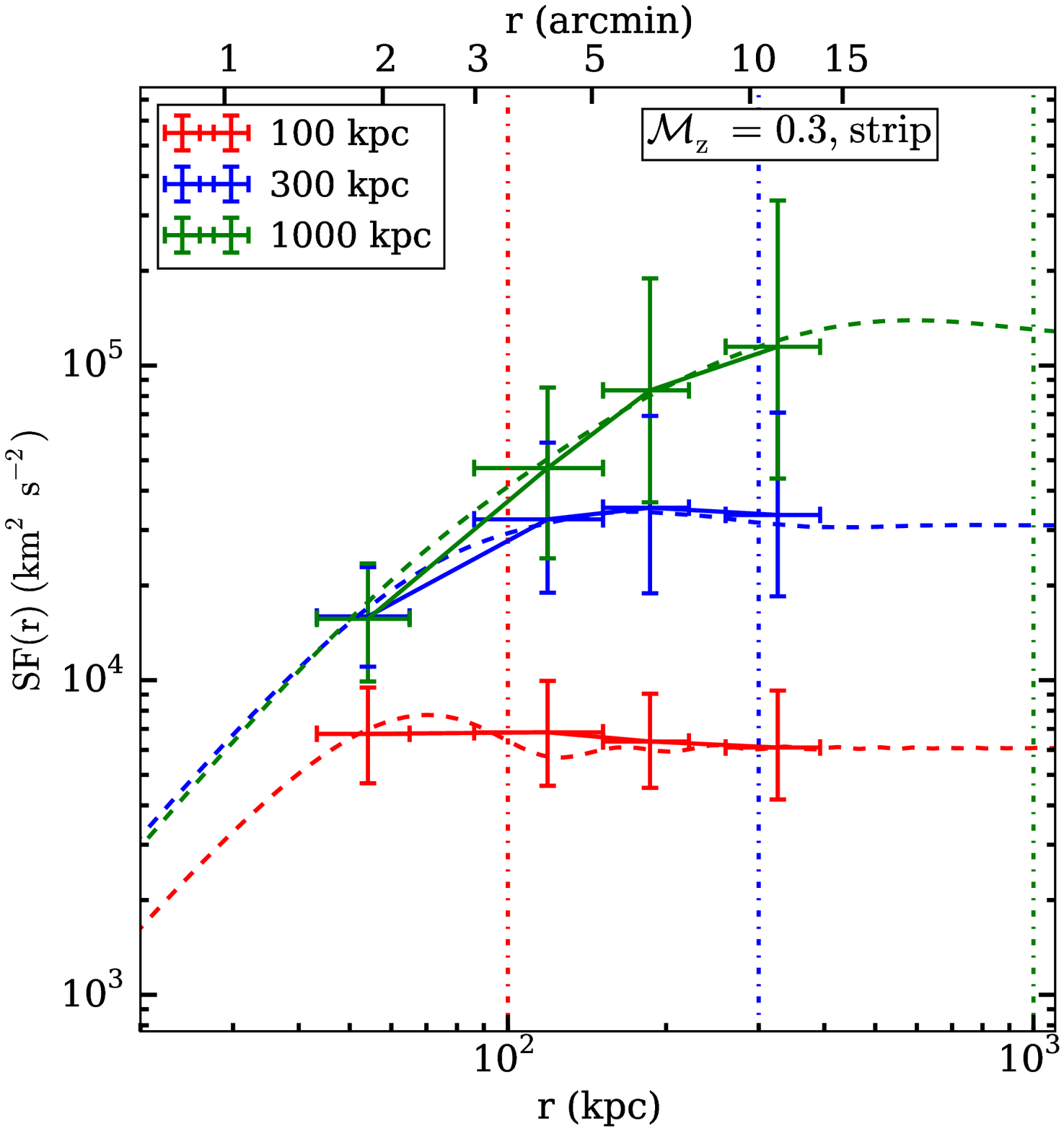}
\includegraphics[width=0.30\textwidth]{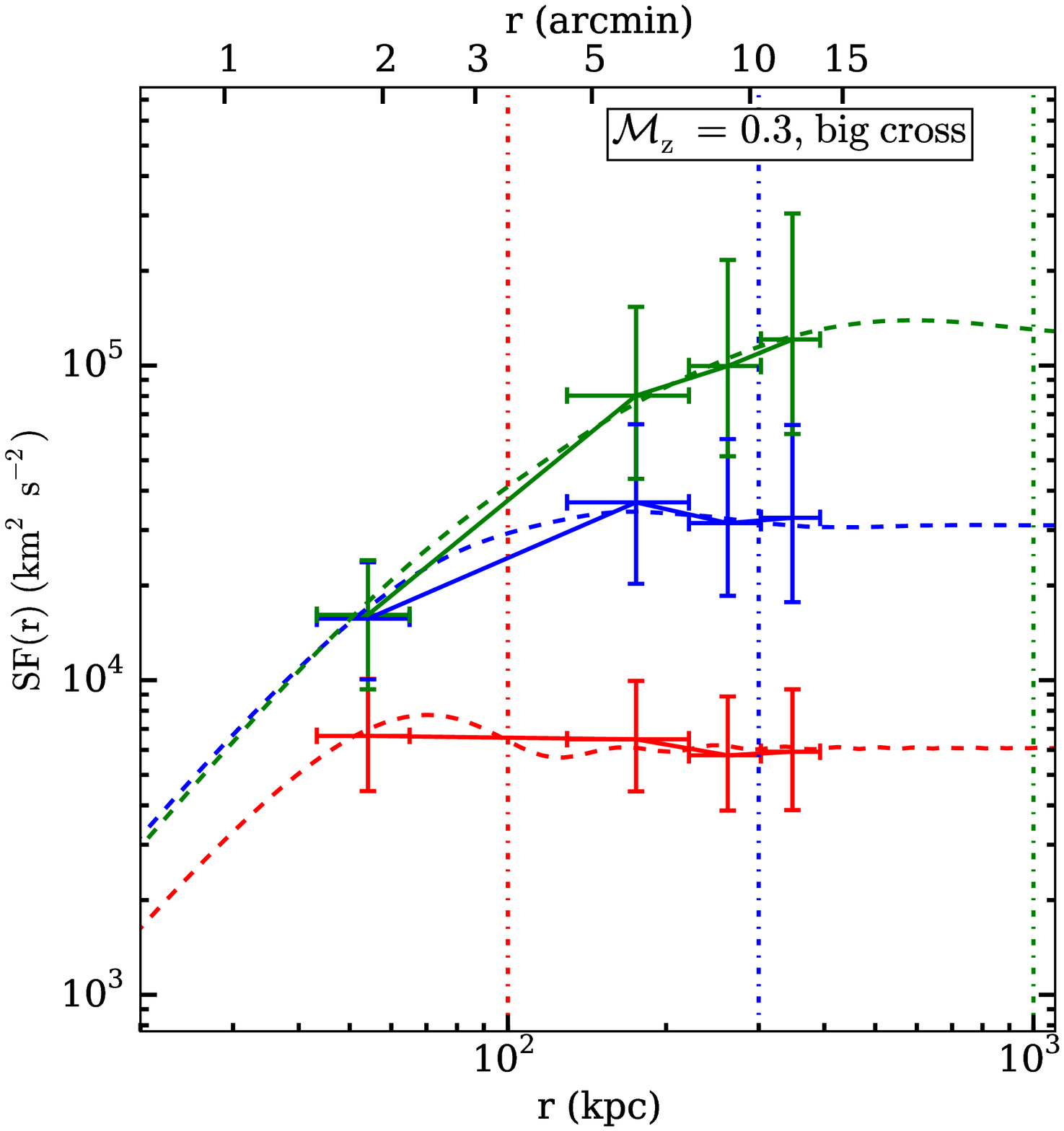}
\includegraphics[width=0.30\textwidth]{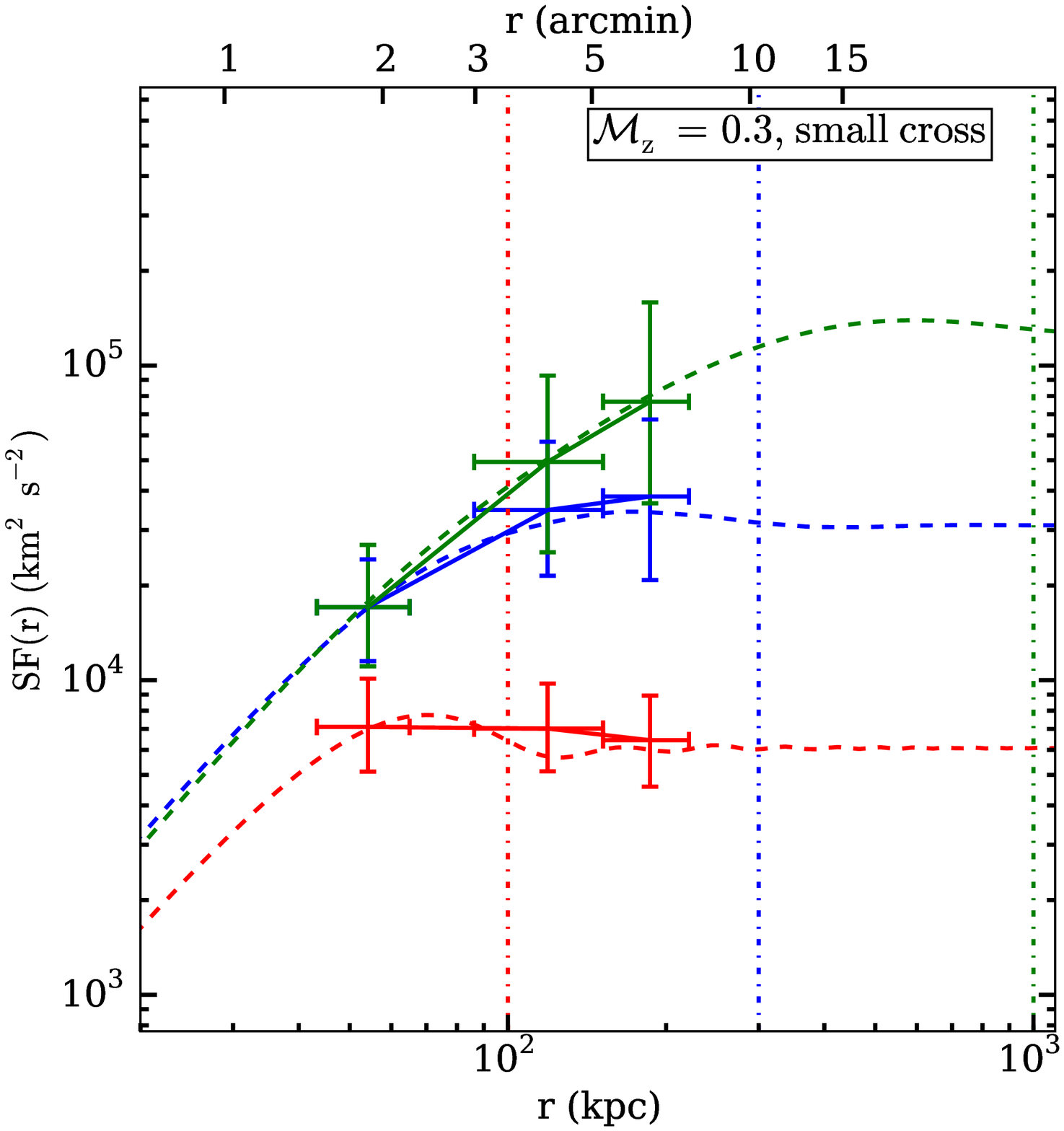}
\includegraphics[width=0.30\textwidth]{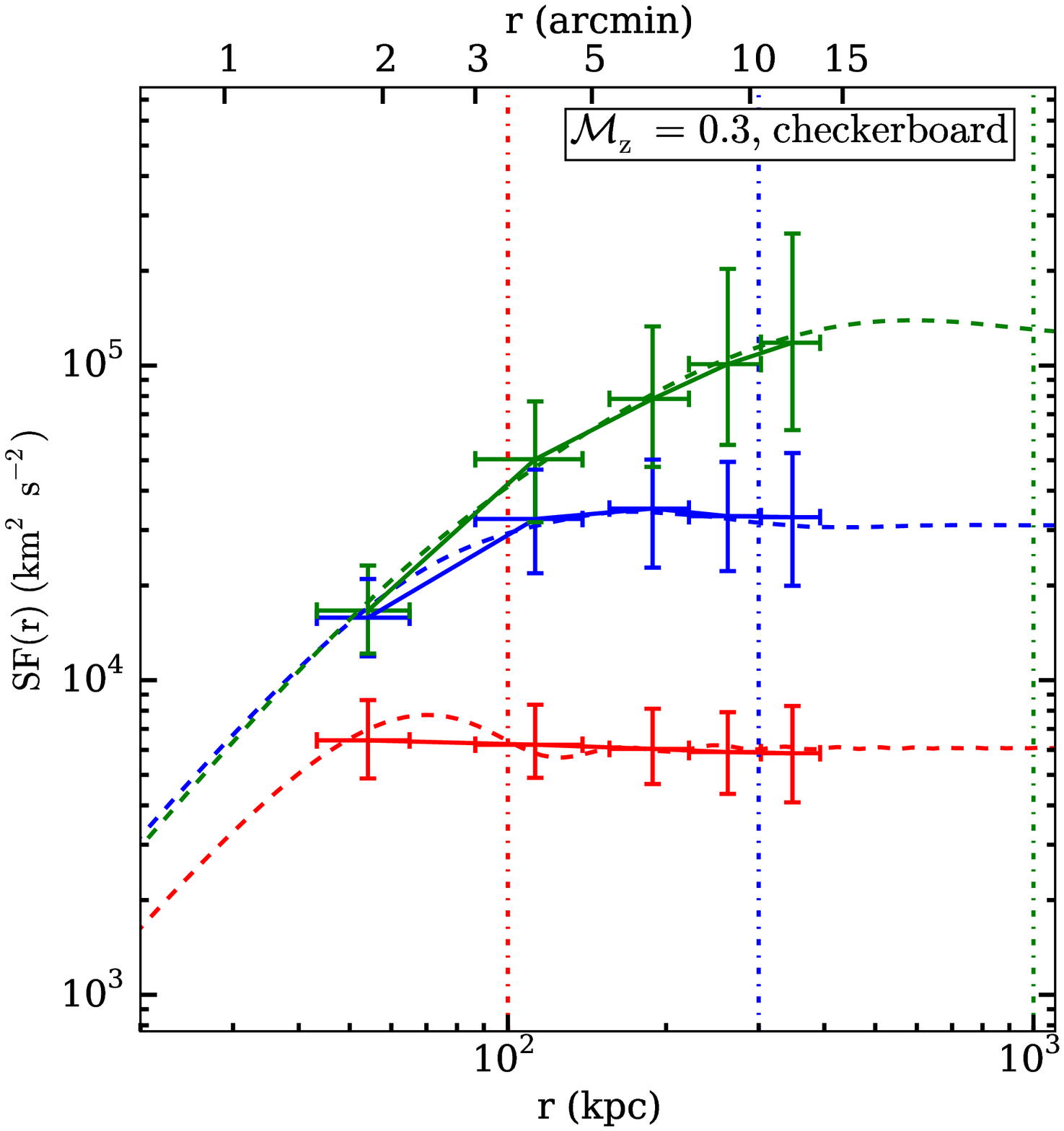}
\includegraphics[width=0.30\textwidth]{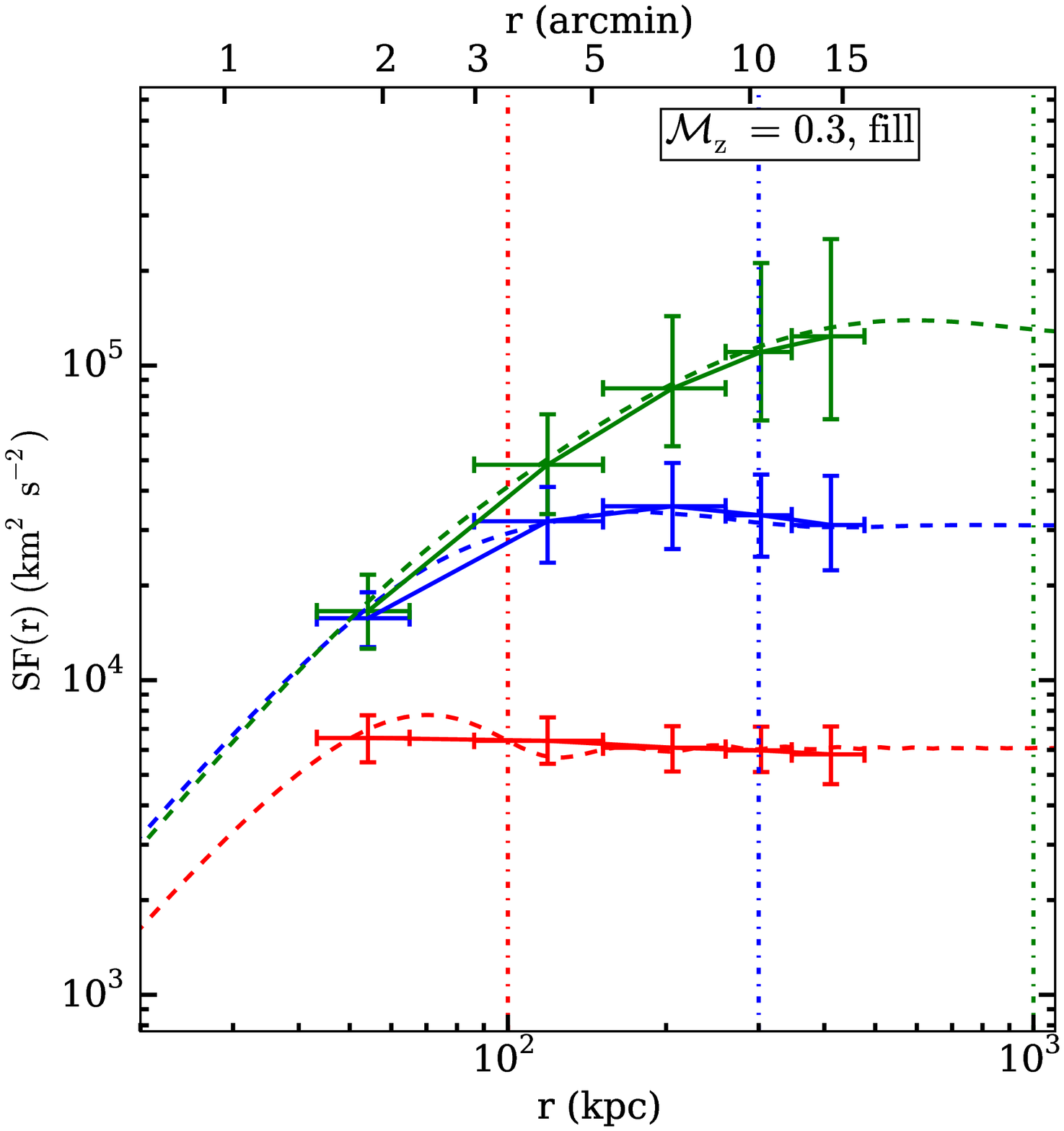}
\caption{Mean and variance of the structure functions for different large-scale cutoffs for ${\cal M}_z = 0.3$ and the five different pointing configurations, computed from 100 realizations of the velocity field. No measurement errors are included, the error bars represent the intrinsic ``cosmic variance'' in the structure function. The analytic prediction is plotted with dashed lines for comparison.}
\label{fig:cosmic_variance}
\end{center}
\end{figure*}

\begin{figure*}
\begin{center}
\includegraphics[width=0.98\textwidth]{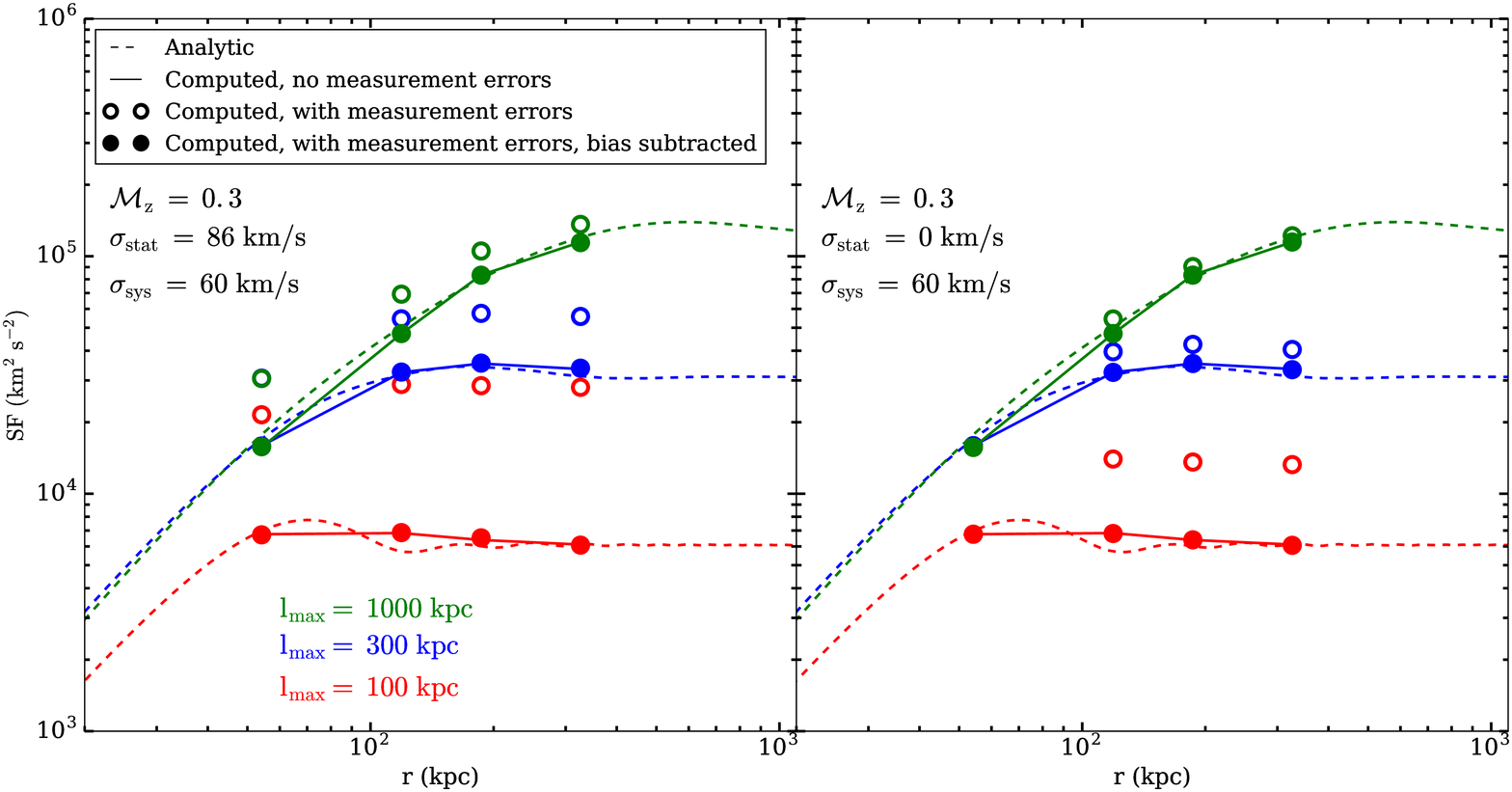}
\caption{The effect of bias on the computed values of the structure function, for different large-scale cutoffs for ${\cal M}_z = 0.3$ and the ``strip'' pointing configuration. Dashed lines indicate the analytic expectation. Solid lines indicate the computed structure function without measurement errors added. Open circles indicate the computed function with the bias due to measurement errors included, and filled circles indicate the computed function with this bias subtracted. Left panel: Bias correction for non-zero statistical error. Right panel: bias correction for zero statistical error (in this case, there is no bias at the smallest scale, since the velocity differences at this scale come from within the same pointing, and there is no systematic error between velocity measurements at this scale, see also Appendix \ref{sec:bias}.)}
\label{fig:error_correction}
\end{center}
\end{figure*}

A given realization of the velocity field in the $z$-direction is generated by setting up a Gaussian random field \citep[we follow a procedure similar to that found in][among other works]{mur04,rus07,rus10,zuh11}. We begin by computing the Fourier transform of the 3D velocity field, $\tilde{v}_z({\bf k})$ = $ve^{i\phi}$, where the amplitude $v$ and the phase $\phi$ for all grid points in $k$-space are drawn randomly from a Rayleigh distribution:
\begin{equation}
P(v,\phi)dvd\phi = \frac{v}{{\Sigma_v^2}}\exp\left(-\frac{v^2}{2\Sigma_v^2}\right)dv\frac{d\phi}{2\pi}
\end{equation}
where $\Sigma_v^2 = P_{\rm 3D}(k)/2$. The normalization $C_n$ of the power spectrum is set by requiring that the line of sight velocity dispersion $\langle{w_z^2}\rangle$ in the cluster averaged over a number of realizations is equal to a fiducial value given by $w_{z,0} = {\cal M}_zc_s$, where ${\cal M}_z$ is the Mach number of the velocity dispersion, which will be varied for different realizations of the structure function, and $c_s$ is the sound speed of the thermal gas. This value is related to the power spectrum by (Z12, Equation 5):
\begin{equation}\label{eqn:avg_vel_disp}
\langle{w_z^2}\rangle = \displaystyle\int{P_{\rm 3D}}({\bf k})[1-P_\epsilon(k_z)]d^3{\bf k}
\end{equation}

The inverse Fourier transform of $\tilde{v}_z({\bf k})$ yields the 3D velocity field $v_z({\bf x})$. This field is then projected along the line of sight in the manner of equations \ref{eqn:los_vel2} and \ref{eqn:los_disp} to produce 2D maps of the line of sight velocity (the line centroid) $\bar{v}_z(x,y)$ and standard deviation (the line width) $w_z(x,y)$. These maps are then reblocked by a factor of 4, resulting in pixels of width 1.5' ($\approx$~41.4~kpc), which is approximately the smallest angular scale that SXS will be able to resolve. We experimented with other options for reblocking the image, and found that this value represents the best compromise between reducing the statistical errors on the line shift and width and resolving power on small scales adequately. We did not include the effect of PSF scattering (recall for SXS, the PSF FWHM $\sim$1'); it will be mild for our chosen pixel size and the flat brightness distribution of a Coma-like cluster, and will be easily accounted for in the analysis of the real data.

In our analysis, the relevant quantities related to the velocity field will be computed within several different configurations of multiple pointings of width 3' (the SXS field of view). They consist of the following (seen in Figure \ref{fig:pattern_shapes}): a single ``strip'' shape 5 pointings (15') across, a ``big cross'' shape with the pointings spaced 6' apart, a ``small cross'' shape with the pointings spaced 3' apart, a ``checkerboard'' configuration that adds four pointings at the corners of the central pointing to the ``big cross'' configuration, and a ``fill'' configuration that nearly fills the space of the simulated data. In the ``checkerboard'' configuration, we will assume the same total exposure time summed across the pointings as for the ``big cross'' configuration, resulting in a reduction in exposure time per pointing of roughly half. For the ``fill'' configuration, we will assume the same exposure {\it per SXS pointing} as the ``strip'' and ``big cross'' configurations, resulting in a total exposure 4.2 times larger. All of these configurations fit within the core radius of Coma, where the density is still relatively flat and the effects of stratification will be minor. For this reason, we have not explored pointing configurations with baselines larger than these. Also, the range of baselines these pointing configurations probe represents the most likely range of injection scales from a physical point of view (see above).

From our maps of the line centroid, we compute the structure function by taking velocity differences between pairs of 1.5'$\times$1.5' pixels and averaging these differences within bins at length scale $r$:
\begin{equation}
SF(r) = \langle{s_{ij}}\rangle_r = \langle(v_i-v_j)^2\rangle_r
\end{equation}
The discrete nature of our maps results in a small number of unique distance scale values between pixels, which suggests a simple binning scheme where similar length scales are grouped together. We have experimented with the number and position of bins and have found that our choice of 4-5 bins per pointing configuration, determined by the similarity of length scales, is adequate to resolve the shape of the structure function.

\subsection{Characterization of Sources of Error}\label{sec:errors}

In this section, we describe the sources of error affecting the measurement of the second-order structure function from line shifts measured by {\it Astro-H}. In the discussion that follows, we only present examples where the injection scale of the turbulent motions is varied, since the variances between these different velocity fields are more pronounced than when the dissipation scale of the turbulence is varied.

\input tab1.tex

\subsection{Cosmic Variance}\label{sec:cosmic_variance}

The structure function $SF(r)$ (and the equivalent power spectrum $P(k)$) represents an average over all possible random realizations of the underlying velocity distribution function. For a given realization of a velocity field, a structure function may be computed from the velocity differences over all length scales within the field, but this will be only a single realization of the underlying distribution. A real cluster in the sky will similarly produce just a single realization of the structure function. To represent this ``cosmic variance,'' for each of the different models of the underlying power spectrum, we generate 100 realizations of the velocity field. For the $k$-th velocity field, we compute the structure function $SF_k(r)$ within the chosen pointing configuration, and then compute the mean and variance of all the $SF_k(r)$.

Figure \ref{fig:cosmic_variance} shows an example computation of the mean and 1-$\sigma$ variance of the structure function calculated for three values of the injection scale $l_{\rm max}$, with all other parameters held fixed. No measurement errors are included at this stage, only those from ``cosmic variance.'' This suggests that distinguishing between different underlying models will be limited by this variance. As we will see, the importance of the cosmic variance is different for different choices of the coverage of velocity measurements from the cluster emission. Needless to say, with more pointings, the cosmic variance errors are reduced. This is especially true in the ``fill'' configuration. However, this configuration would require a large and expensive survery. Below, we will look for a compromise between the need to sample more sight lines and the need to limit the total required exposure.

\subsection{Measurement Errors}\label{sec:measurement_errors}

Once a given realization of the velocity field is generated, the effect of statistical and systematic errors on the measurement of the line shift and width must also be taken into account. We simulate statistical errors on the line shift by adding a normally distributed velocity component with standard deviation $\sigma_{\rm stat}$ to each pixel, which depends on the line width, or the Mach number of the turbulence. We also estimate the statistical error on the line width (which will be used in Section \ref{sec:estimating_inj_scale} for an additional constraint on the velocity field parameters). For a given line width, the statistical error on the line energy scales as roughly $\sqrt{N_{\rm counts}}$ (where $N_{\rm counts}$ is the number of counts in the line), and also increases for increasing line width, or $w_{z,0}$. Table \ref{tab:stat_err} shows the estimated statistical errors on the line shift and width for a single 1.5'$\times$1.5' pixel for several different values of the line of sight turbulent Mach number ${\cal M}_z$, assuming an exposure time per pointing of $\sim$100~ks (corresponding to roughly 125~counts in the Fe-K line per pixel for a Coma-like cluster).\footnote{The results here are based on the statistical accuracy on the line velocity derived using $\sim$125 line counts. Those counts can come from the He-line Fe line ($E$ = 6.7~keV) alone, or from the He-like and H-like (6.9 keV) lines combined; the latter gives a similar formal accuracy for the same total counts. For the Coma gas temperature, the
He-like to H-like line ratio is about 10:7, so the exposures that correspond to our simulated statistical accuracy would depend on
whether the observer chooses to combine the two lines ($\sim 100$ ks is for the use of the He-like line alone). This will require a
judgment on whether the two lines sample the same gas along the l.o.s. (e.g., there are no projected hot clouds, etc.) based on a more
detailed study of the Coma spectrum. Here we assume that the line emissivity (per unit emission measure) is uniform throughout the
cluster.} Additionally, to take into account a conservative systematic error on the line energy from the systematic uncertainty of SXS gain stability, we add a normally distributed velocity component of $\sigma_{\rm sys}$ = 60~km/s (equivalent to $\sim$1.3~eV at $E \sim 6.5$~keV) to the velocity. The added systematic uncertainty is constant for a given 3'$\times$3' pointing, e.g., for velocity differences within the same pointing, the systematic error cancels out. For each of the 100 realizations of the velocity field, we compute 100 realizations of the measurement error, resulting in a total of 10$^4$ realizations of the velocity field for each combination of the input parameters. These realizations will determine the scatter of the structure function due to both ``cosmic variance'' and ``measurement'' errors.

\begin{figure*}
\begin{center}
\includegraphics[width=0.98\textwidth]{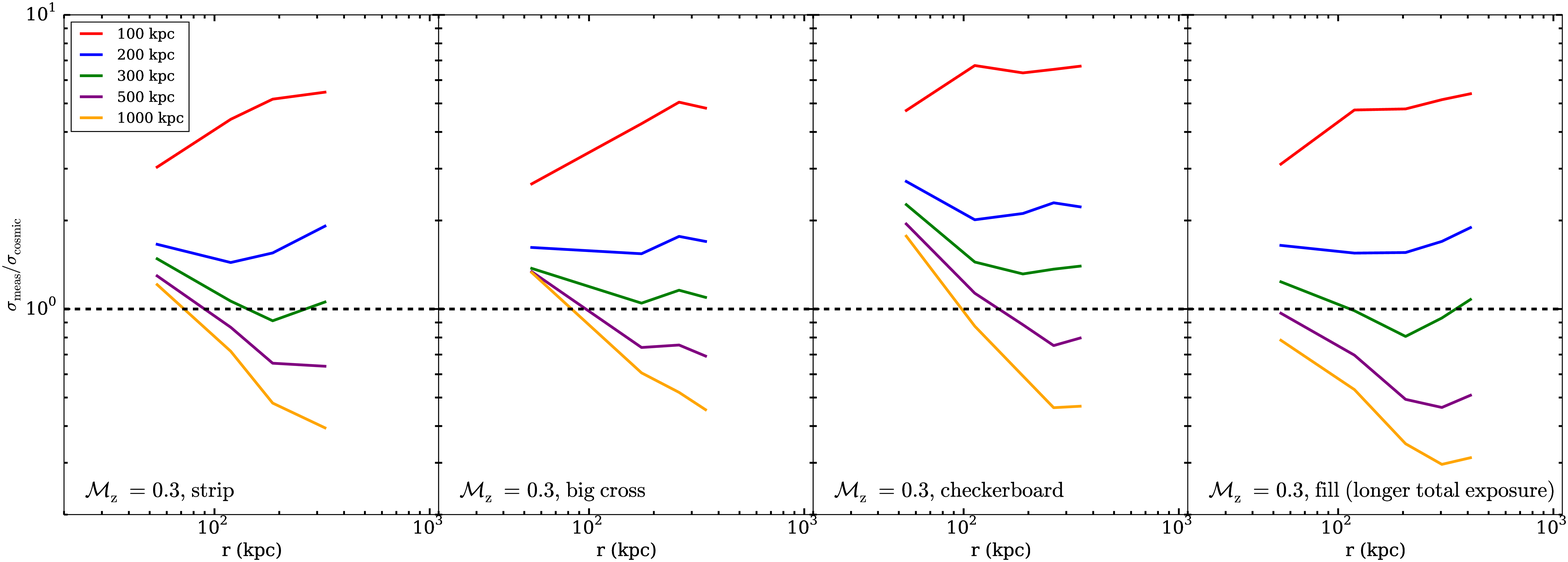}
\caption{The ratio of measurement error to cosmic variance error only for the ``strip,'' ``big cross,'' ``checkerboard,'' and ``fill'' configurations for structure functions with different injection scales. The ``strip,'' ``big cross,'' and ``checkerboard'' configurations all have the same total exposure, whereas the ``fill'' configuration has the same exposure per pointing as ``strip,'' for a longer total exposure.}
\label{fig:meas_vs_cosmic_err}
\end{center}
\end{figure*}

\begin{figure*}[!thp]
\begin{center}
\includegraphics[width=0.32\textwidth]{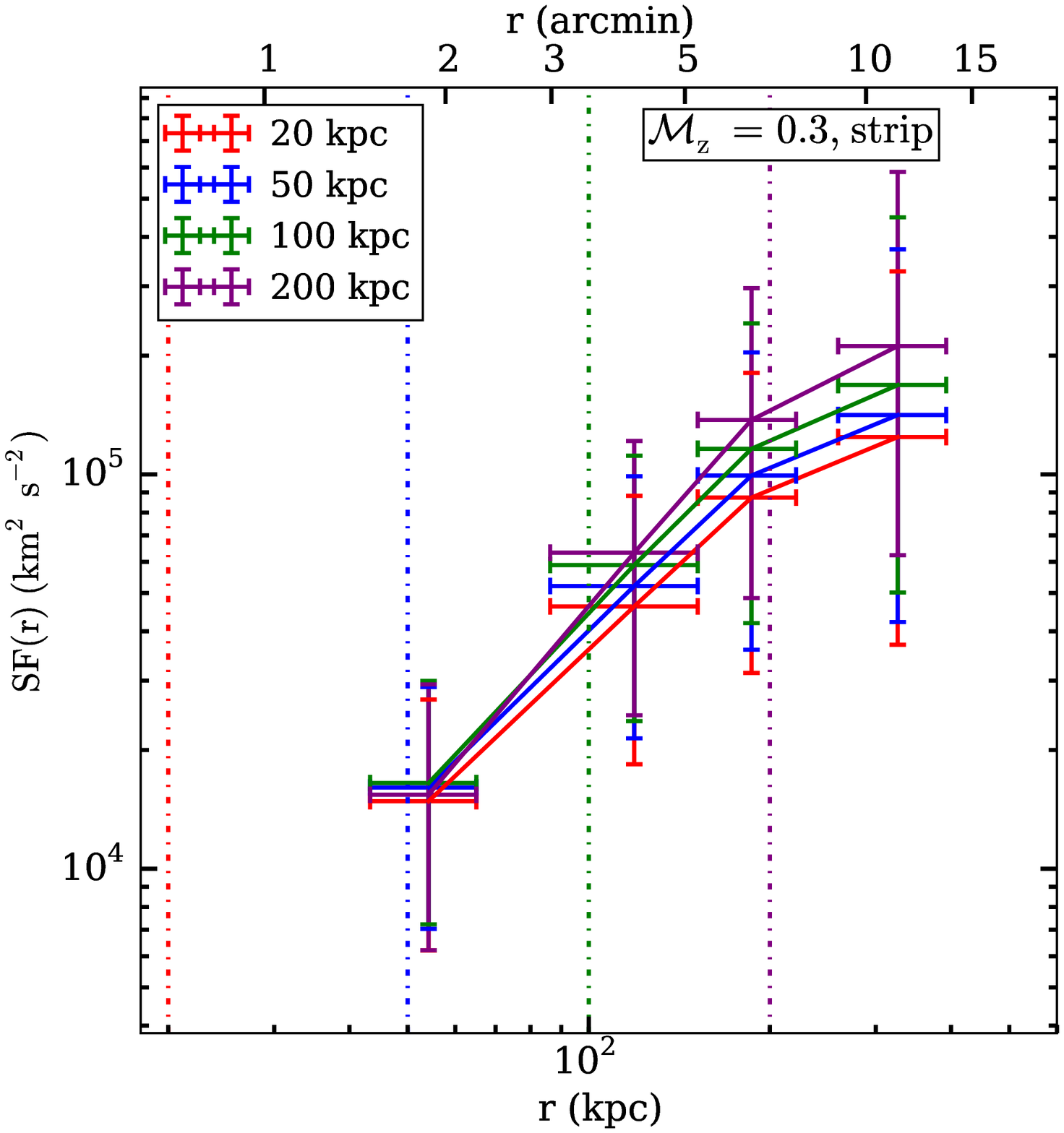}
\includegraphics[width=0.32\textwidth]{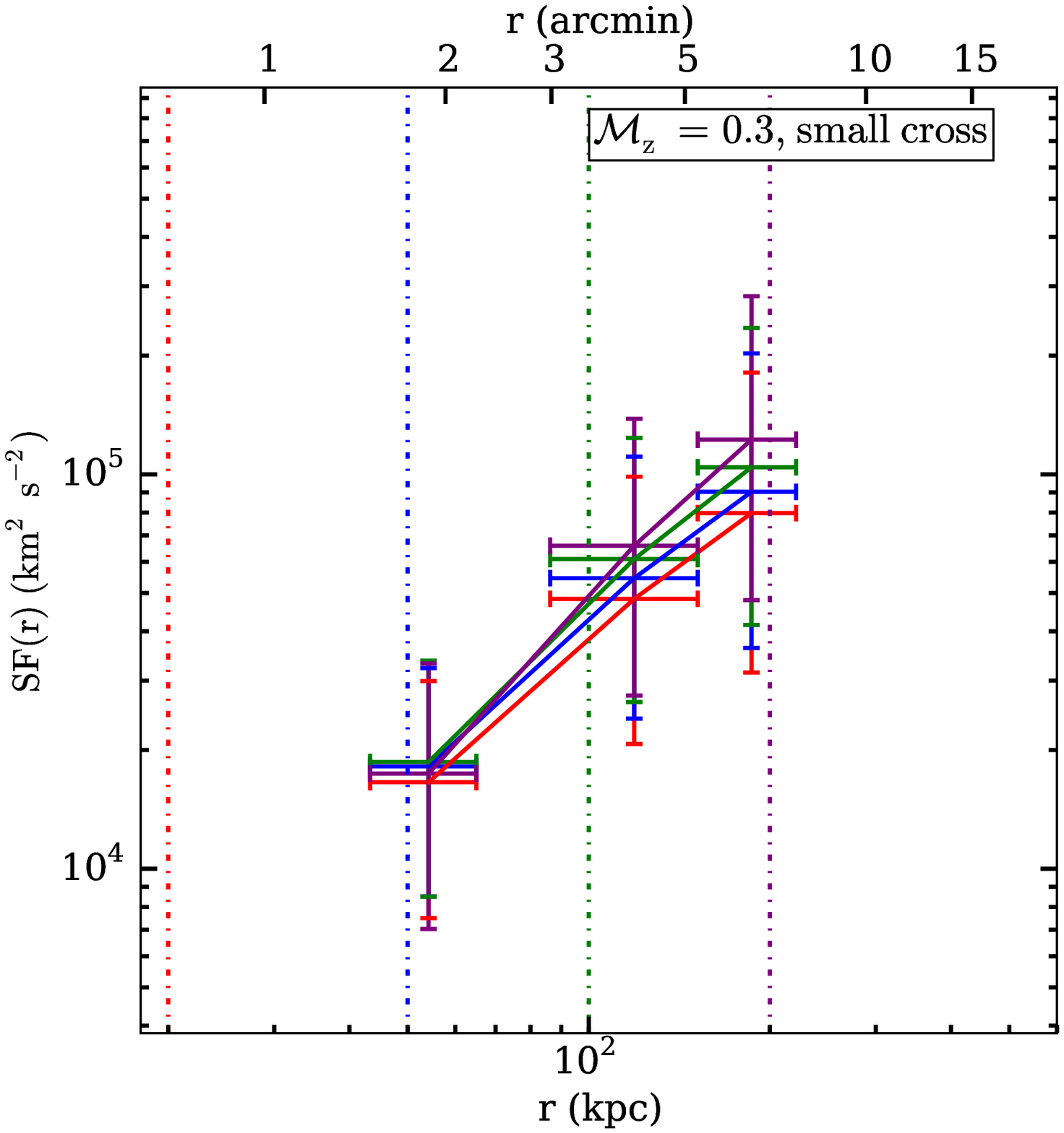}
\includegraphics[width=0.32\textwidth]{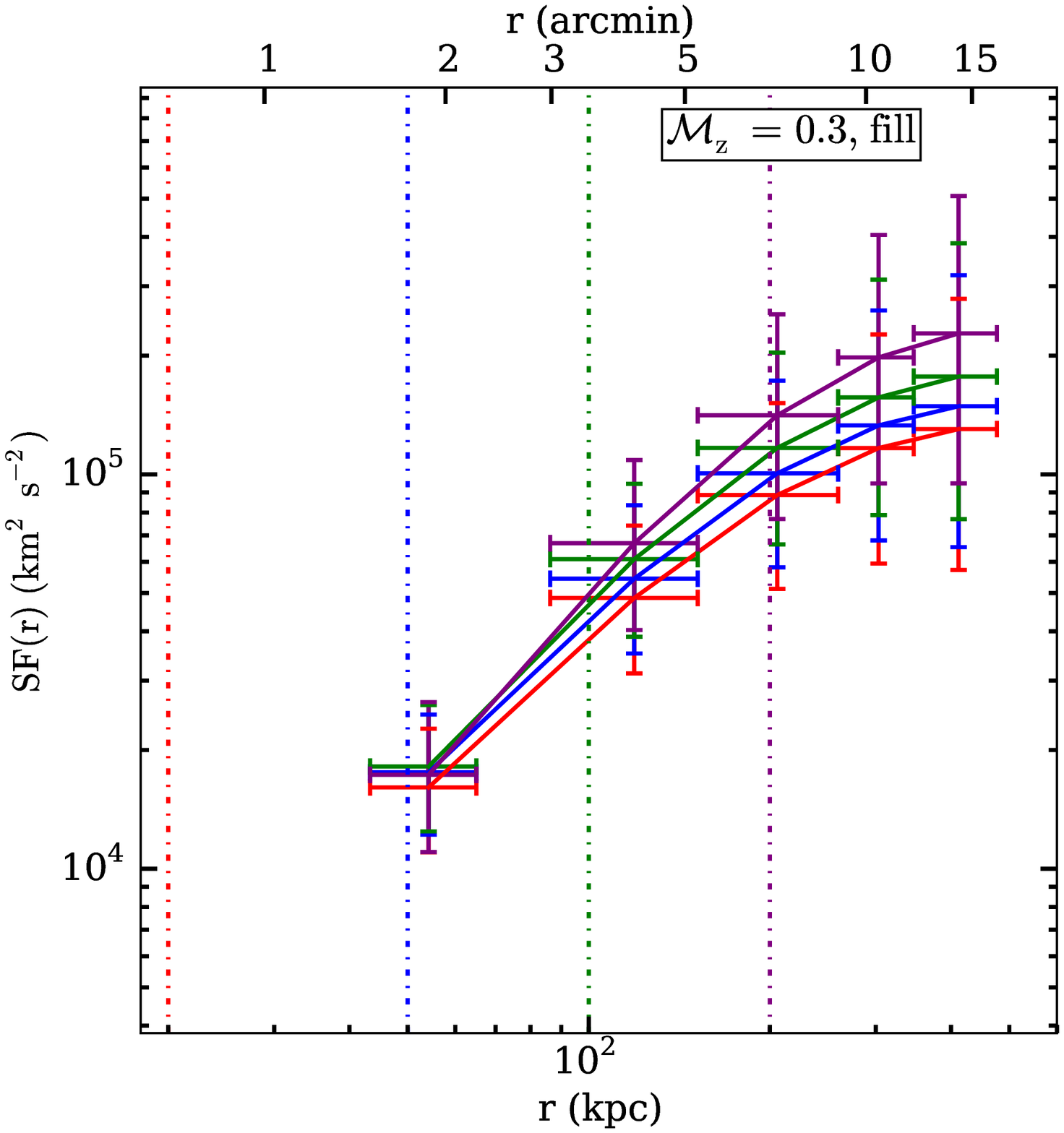}
\caption{Structure functions for input power spectra with different dissipation scales $l_{\rm min}$, for ${\cal M}_z$ = 0.3 and three different pointing configurations. Colored dot-dash lines indicate the locations of the cutoff scales.}
\label{fig:small_cutoffs}
\end{center}
\end{figure*}

The addition of measurement errors necessarily introduces a bias in the computation of the structure function. Since these velocity differences add in quadrature when we compute the structure function, the end result is that the observed structure function $SF'(r)$ is biased upward from the ``true'' value $SF(r)$. Under the assumptions of uniform $\sigma_{\rm stat}$ and $\sigma_{\rm sys}$ per pointing, we can compute the bias to be (see Appendix \ref{sec:bias}):
\begin{equation}\label{eqn:meas_bias}
SF'(r) = SF(r) + 2\sigma_{\rm stat}^2 + 2\sigma_{\rm sys}^2
\end{equation}
where for the smallest length scales (within the 3'$\times$3' SXS field of view), $\sigma_{\rm sys}$ = 0.

\begin{figure*}
\begin{center}
\includegraphics[width=0.45\textwidth]{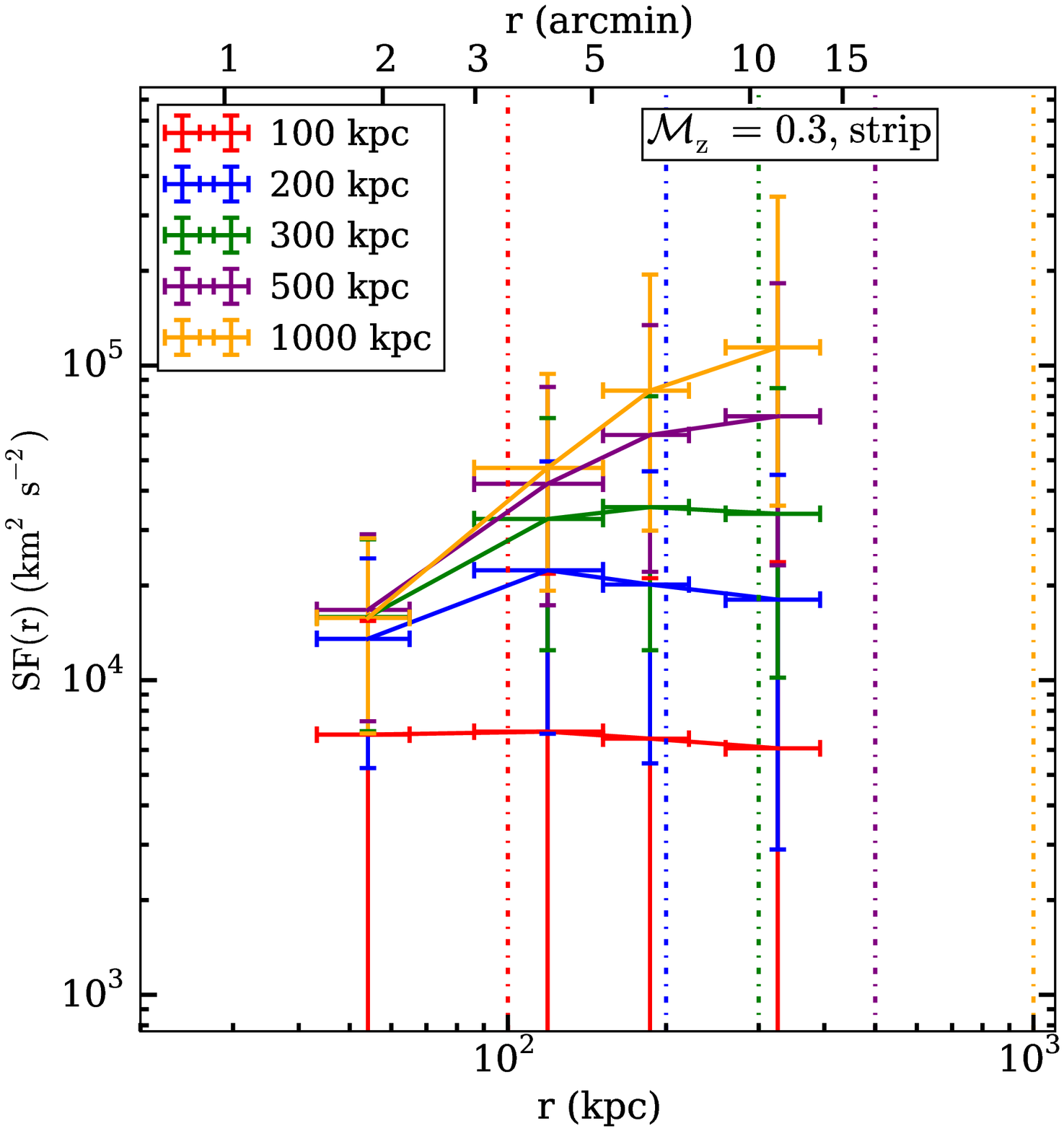}
\includegraphics[width=0.45\textwidth]{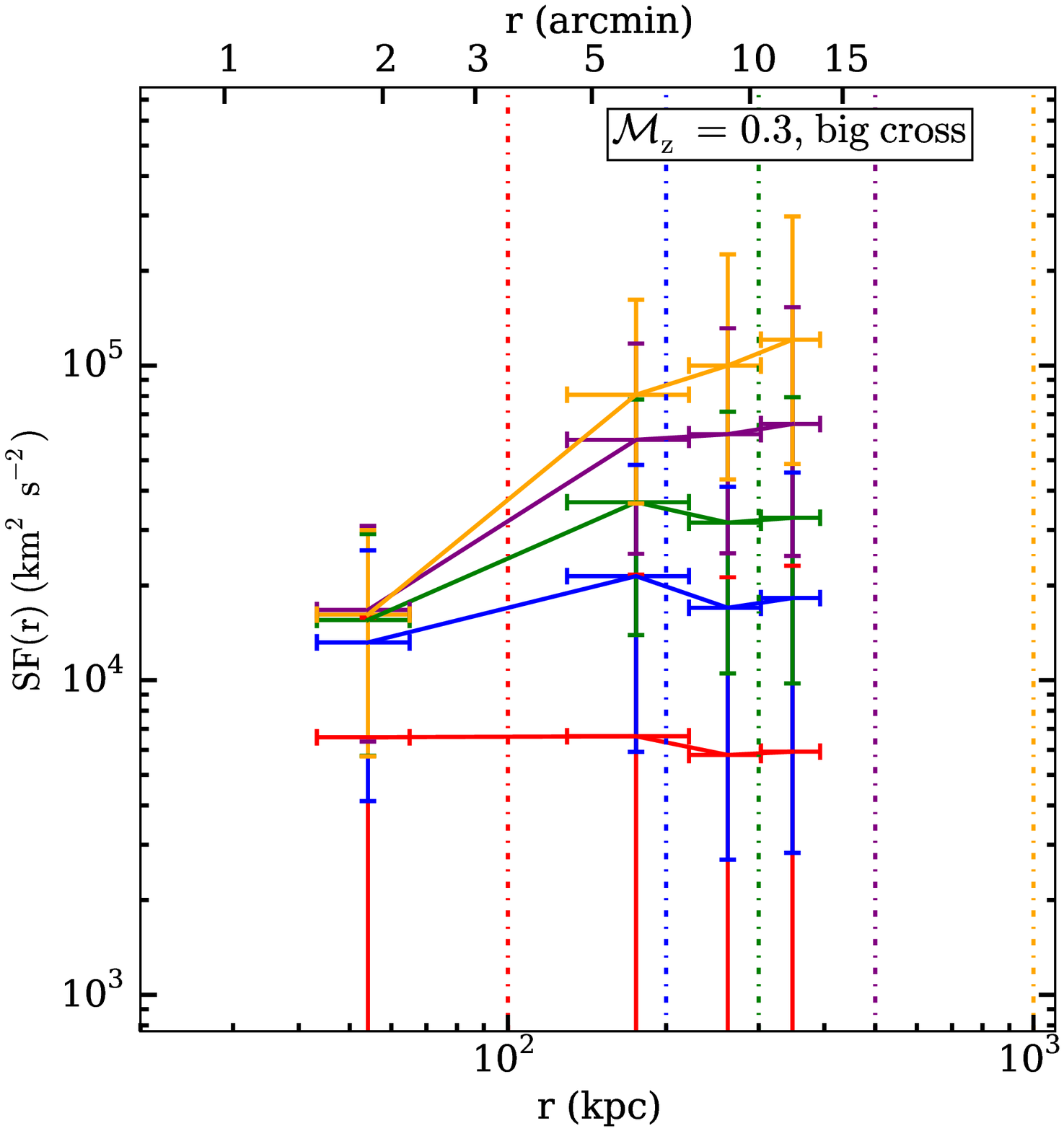}
\includegraphics[width=0.45\textwidth]{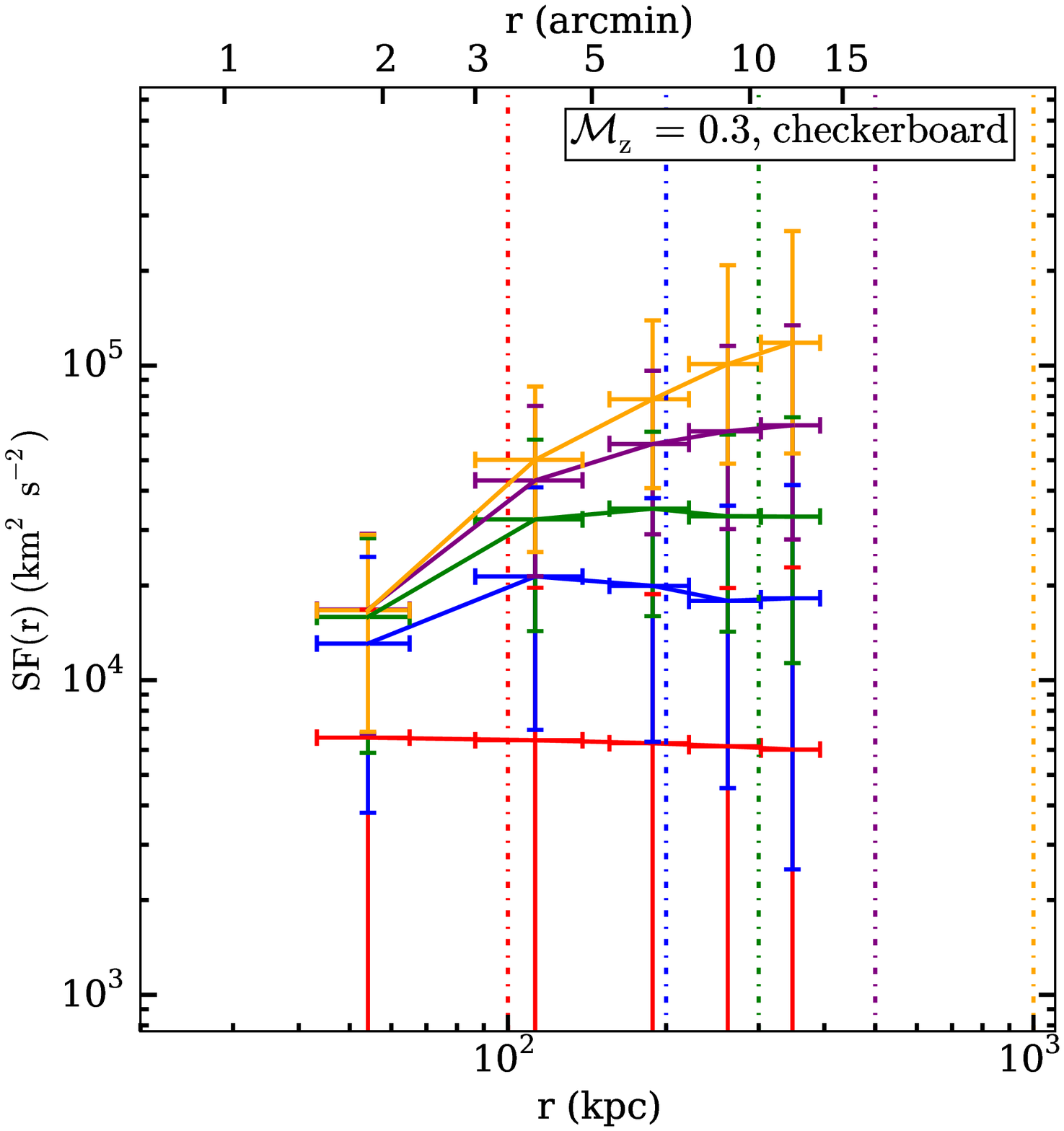}
\includegraphics[width=0.45\textwidth]{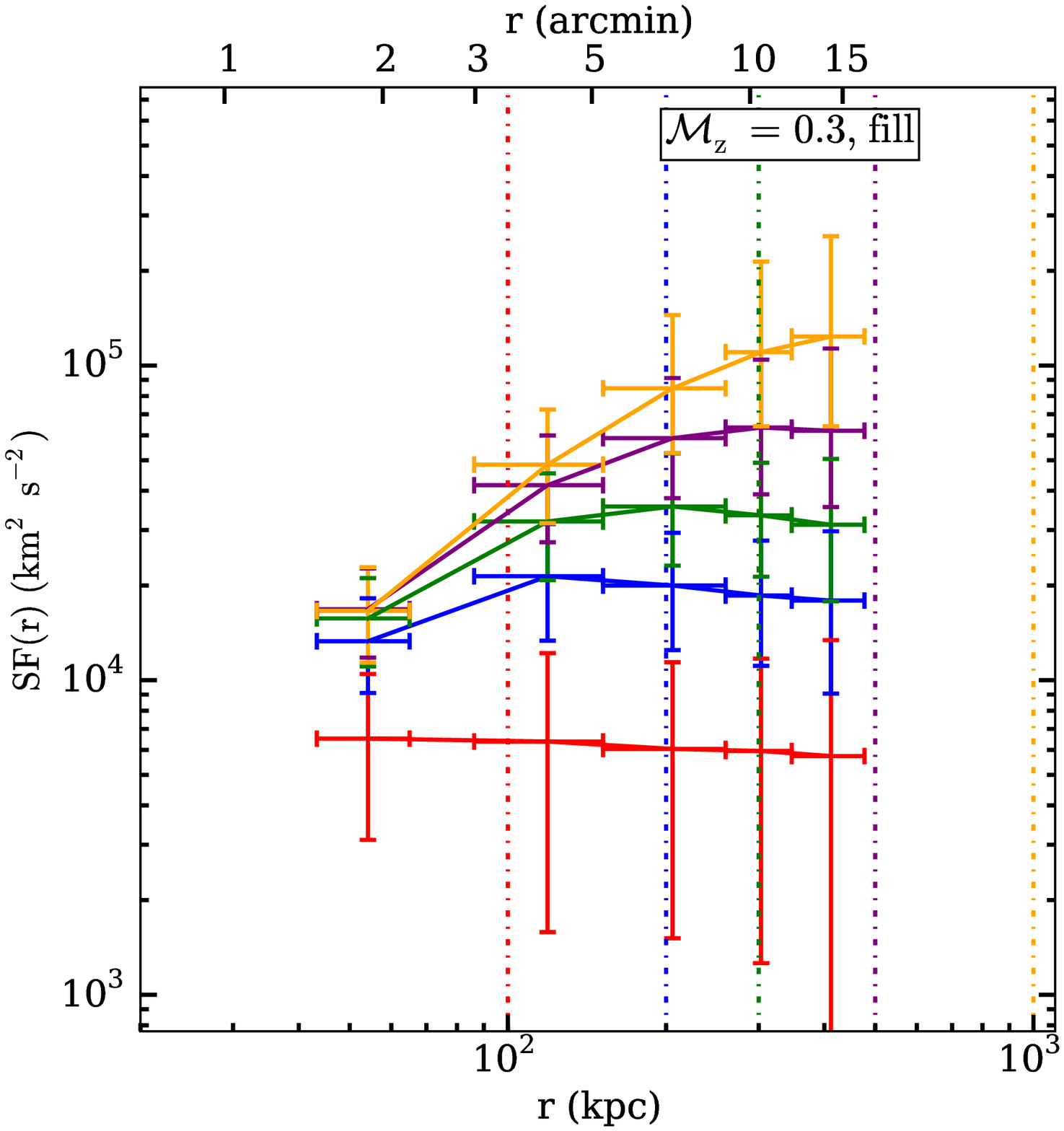}
\caption{Structure functions for input power spectra with different injection scales $l_{\rm max}$, for ${\cal M}_z$ = 0.3 and four different pointing configurations. Colored dot-dash lines indicate the locations of the cutoff scales.}
\label{fig:large_cutoffs}
\end{center}
\end{figure*}

Figure \ref{fig:error_correction} shows the computed structure functions with and without measurement errors, to illustrate the effect of the bias, as well as the structure function with the bias subtracted, following Equation \ref{eqn:meas_bias}. First, as seen in the figure, the bias is most serious for the structure function with a small injection scale $l_{\rm max}$, where the true velocity differences are small, and dominated by the measurement errors. Second, the bias-subtracted curve agrees well with the analytic expectation, as well as the computed curve generated without measurement errors. We present two cases, one where the statistical error is set to the expected value for our selected exposure time, and another with vanishing statistical error, in the limit of large exposure time (so only the systematic errors, which are independent of exposure time, are included). We see from this comparison that the dominant contribution to the bias on the structure function is from the statistical error. All of our results from this point on are presented with this bias (both statistical and systematic) corrected.

The measurement errors will increase the variance of the structure function beyond that expected from ``cosmic variance'' in the velocity field. Figure \ref{fig:meas_vs_cosmic_err} shows the ratio of measurement and cosmic variance error for different pointing configurations and different injection scales. The velocity field with an injection scale of 100~kpc is most severely affected by the measurement errors, so much so that they dominate the total error. For the other injection scales, the measurement error is typically comparable to the cosmic variance error, for our particular choice of exposures. In particular, for the ``strip,'' ``big cross,'' and ``checkerboard'' configurations, which divide the same total exposure differently among several offsets, these errors are close for the mid-range value of the injection scale, which means that the exposures are near-optimal.

\section{Results}\label{sec:results}

\subsection{Small-scale Cutoffs/Dissipation Scales}\label{sec:small_scale}

Figure \ref{fig:small_cutoffs} shows the computed structure functions for small-scale cutoffs of 20, 50, 100, and 200~kpc for the ``strip,'' ``small cross,'' and ``fill'' pointing configurations and for a value of ${\cal M}_z$ = 0.3, all assuming an exposure time of $\sim$100~ks per pointing (so ``fill'' has 4.2 times the total exposure of the other two configurations). In each of the figures, the vertical error bars correspond to the 1-$\sigma$ errors (including both the cosmic variance and measurement errors) and the horizontal error bars delineate the bin sizes. In all of the pointing configurations, all of these curves are indistinguishable within the 1-$\sigma$ errors. In addition, the scales below $\sim$40~kpc are not resolved by the 1.5' pixels. Unless the dissipation scale is much larger than 200~kpc (an unexpected scenario, given the theoretical motivations for the expected range of this parameter; see Section \ref{sec:defining_params}), {\it Astro-H} will not be able to distinguish the corresponding structure function from one with a smaller dissipation scale.

\subsection{Large-scale Cutoffs/Injection Scales}\label{sec:large_scale}

Figure \ref{fig:large_cutoffs} shows the computed structure functions for large-scale cutoffs of 100, 200, 300, 500, and 1000~kpc for the different pointing configurations, for a value of ${\cal M}_z$ = 0.3, assuming a $\sim$500~ks total exposure for each configuration, with the exception of the ``fill'' configuration, which has a total exposure of $\sim$2100~ks. This shows that the possibility exists for distinguishing between structure functions with different injection scales, due to the very different shapes of the curves. The sensitivity to the injection scale depends on the scale of interest and the configuration of the pointings.

Distinguishing between the curves at small separations ($r \simlt 100$~kpc) is difficult due to the proximity of the curves, regardless of the pointing configuration. The curves will be most easily distinguished at large spatial scales $r$. At these scales, the ``big cross'' configuration (top-right panel) allows for the differences in the structure functions to be better constrained than the ``strip'' configuration, due to the increased number of pairs of points at large distances. On the other hand, the ``strip'' pointing allows for the shape of the structure function in the intermediate range of $r$ ($\sim$80-150~kpc, 3-5') to be better constrained, due to the availability of distance measurements within this range for this pointing. This is the range of length scales where the structure function from velocity fields with $l_{\rm max}$ = 200 and 300~kpc transitions from a power-law dependence on $r$ to a constant value.

The ``fill'' configuration (lower-right panel) combines both of the strengths of these two pointings, and improves the resolving power at all scales, making it much easier to distinguish between the curves. However, it requires 4.2 times the exposure of the other configurations. The ``checkerboard'' configuration represents a compromise between the ``big cross'' and ``fill'' configurations, by adding four pointings at the corners of the central pointing to increase the number of measurements at small-to-intermediate scales. Here, we have assumed that the {\it total} exposure time (summed over the pointings) for the ``checkerboard'' is the same as the total for ``big cross'' ($\sim$500~ks), so that each individual 3'$\times$3' pointing has an exposure time of $\sim$55~ks (corresponding to $\sim$70~line counts per 1.5'$\times$1.5' pixel). From the perspective of being able to distinguish between the structure function curves at intermediate to large length scales, ``checkerboard'' is a modest improvement over that of ``big cross'' (see also Section \ref{sec:estimating_inj_scale}). We note that we assume the statistical error on the line shift scales as $\sqrt{N_{\rm counts}}$, at even low values of $N_{\rm counts}$; this will have to be verified using the early (real) SXS data.

For smaller injection scales, the inertial range of the structure function will be difficult to discern due to the inability to resolve length scales below $\sim$40~kpc, the smallest length scale we resolve. The $l_{\rm max}$ = 100~kpc and 200~kpc curves are essentially flat over the entire range of $r$ that is discernable by any of the chosen pointing configurations. Additionally, these curves are most affected by the measurement errors, due to the smaller velocity differences. In particular, the $l_{\rm max}$ = 100~kpc curve is often nearly consistent with zero due to the uncertainty from the combined effect of the statistical and systematic errors on the line shift.

\subsection{Different Power-law Slopes}\label{sec:slopes}

\begin{figure}
\begin{center}
\includegraphics[width=0.45\textwidth]{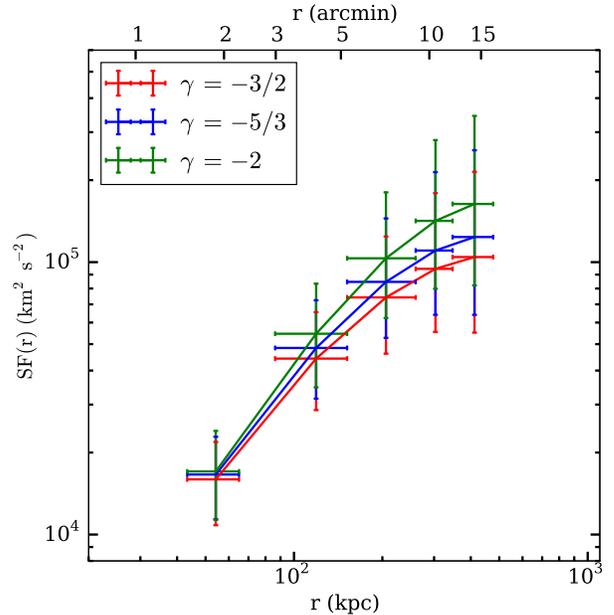}
\caption{Structure functions for input power spectra with different power-law slopes $\alpha$, for ${\cal M}_z = 0.3$, and the ``fill'' pointing configuration. For all the curves, $l_{\rm min}$ = 20~kpc and $l_{\rm max}$ = 1000~kpc.}
\label{fig:different_slopes}
\end{center}
\end{figure}

It is of interest to determine the slope of the structure function $\alpha$ in the inertial range (provided this range can be discerned; that is, there is enough of a dynamic range between the injection and dissipation scales), since this is directly related to the power-law slope of the underlying velocity power spectrum, the value of which depends on the physics of the turbulent medium. Figure \ref{fig:different_slopes} shows a comparison of structure functions for three physically motivated values for the slope $\alpha$ discussed in Section \ref{sec:defining_params}, with the cutoff values of $l_{\rm min} = 20$~kpc and $l_{\rm max} = 1000$~kpc held fixed for the different input power spectra, assuming $\sim$100~ks exposure per pointing in the ``fill'' configuration. Similarly to the situation with the small-scale cutoffs, the curves are too similar to be distinguished within the errors. Despite the qualitatively different physics that produces the different power spectra, the associated range of spectral indices is not wide enough to be discerned by this technique.

\begin{figure*}
\begin{center}
\includegraphics[width=0.97\textwidth]{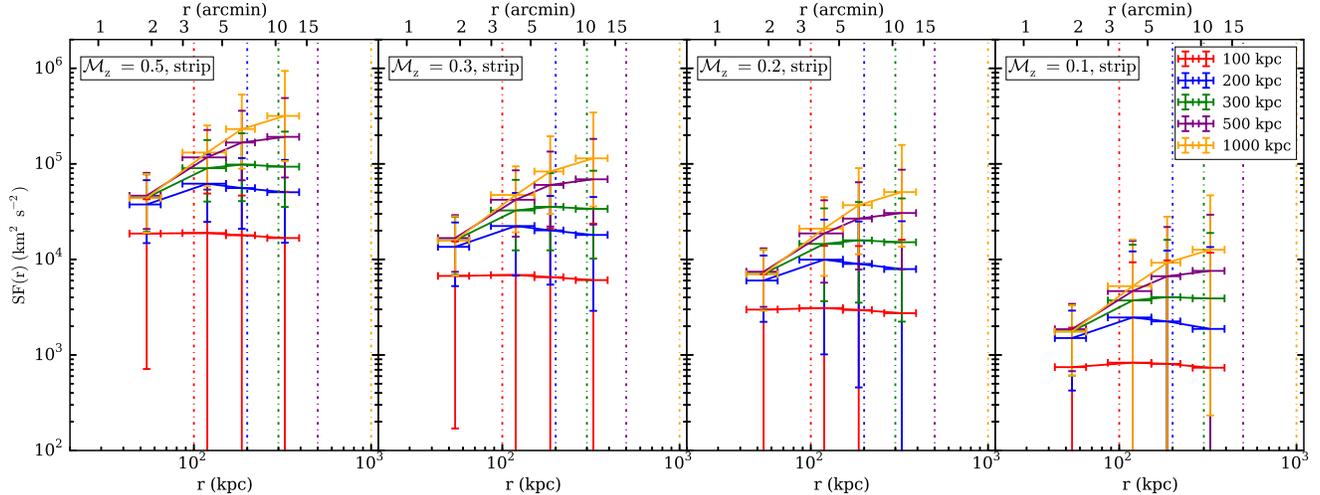}
\caption{Structure functions for input power spectra with different injection scales $l_{\rm max}$, for different Mach numbers ${\cal M}_z$ and the ``strip'' configuration. Colored dot-dash lines indicate the locations of the cutoff scales.}
\label{fig:diff_mach_numbers}
\end{center}
\end{figure*}

\subsection{Different Mach Numbers}\label{sec:mach}

The broadening of emission lines will provide a measure of the strength of the turbulence in the velocity component along the line of sight. An increase in the velocity scale of the turbulence will correspond to an increase in the overall normalization of the power spectrum as well as the measured structure function. Stronger turbulence also implies higher velocities, which will make it easier to measure velocity differences.

On the other hand, for a constant number of counts in the line, the statistical error on the line shift will be roughly proportional to the measured width of the emission line, which contains contributions from thermal, instrumental, and turbulent broadening:
\begin{equation}
w_{\rm obs}^2 = w_{\rm inst}^2+w_{\rm therm}^2+w_{\rm turb}^2
\end{equation}
In our situation, the first two terms will be roughly constant, implying for low $w_{\rm turb}$ the error on the line shift will be approximately constant, and increase for stronger turbulence. The systematic error on the line shift is instrumental in nature and will be independent of the turbulent velocity strength.

Figure \ref{fig:diff_mach_numbers} shows the observed structure functions for different values of the line of sight Mach number ${\cal M}_z$, assuming $\sim$100~ks exposure per pointing. As the overall strength of the turbulence decreases, it will be more difficult to distinguish velocity fields with different injection scales from each other, due to the fact that as the magnitude of the velocity difference decreases, the measurement errors on the line shift are roughly constant. For turbulent Mach numbers ${\cal M}_z < 0.2$ in Coma, we are unlikely to be able to constrain the injection scale of the turbulence.

\begin{figure*}
\begin{center}
\includegraphics[width=0.49\textwidth]{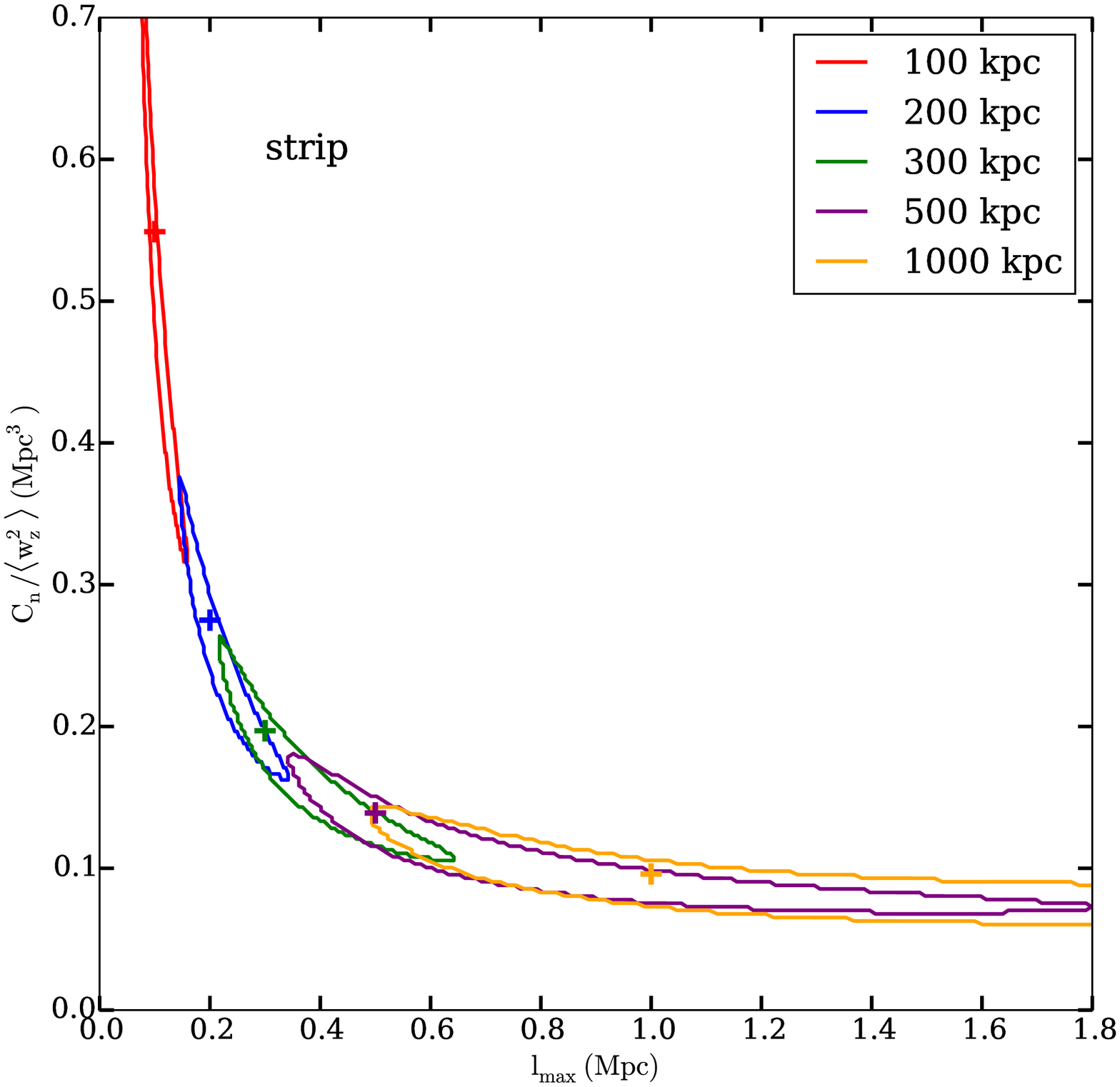}
\includegraphics[width=0.49\textwidth]{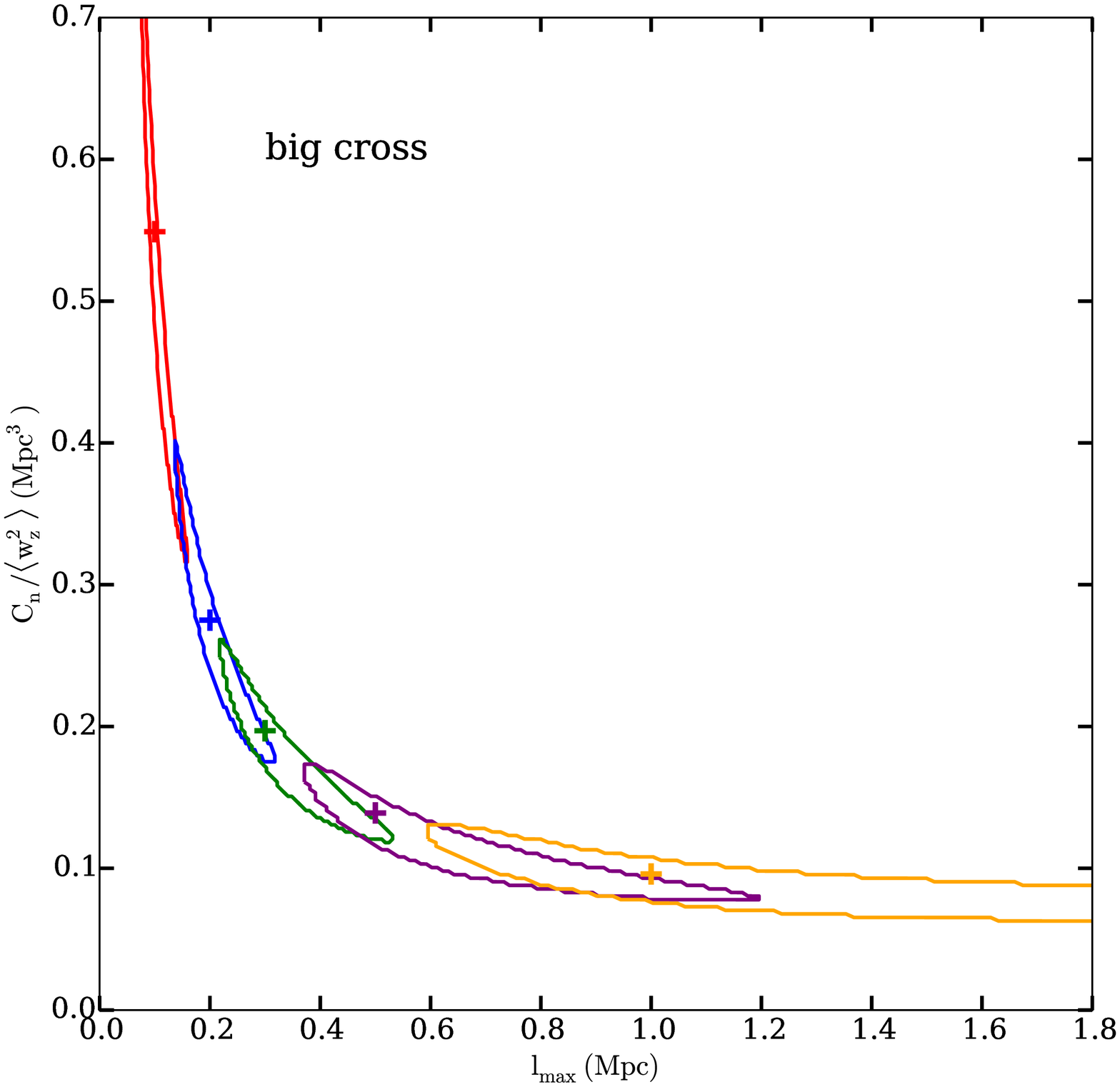}
\includegraphics[width=0.49\textwidth]{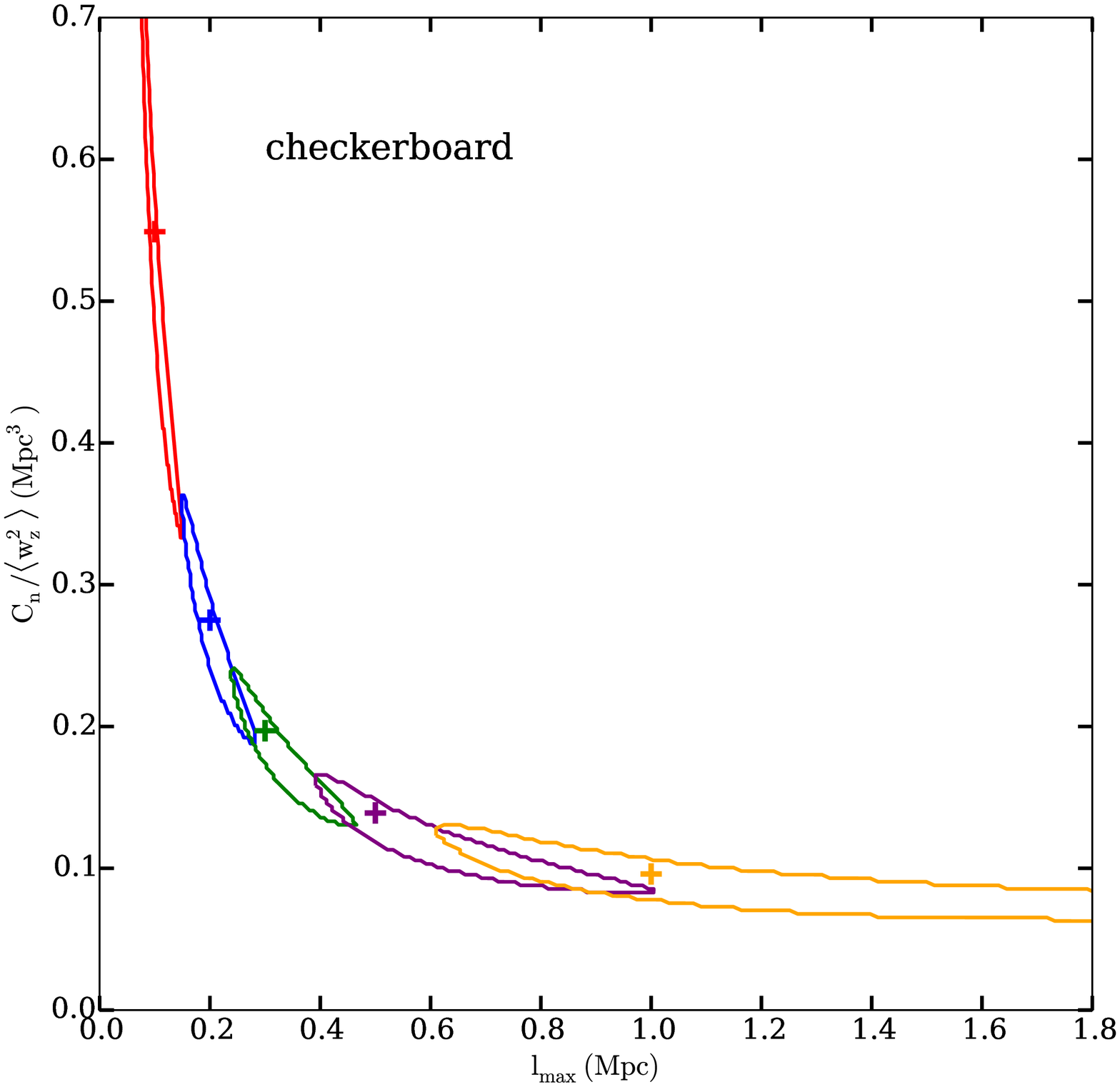}
\includegraphics[width=0.49\textwidth]{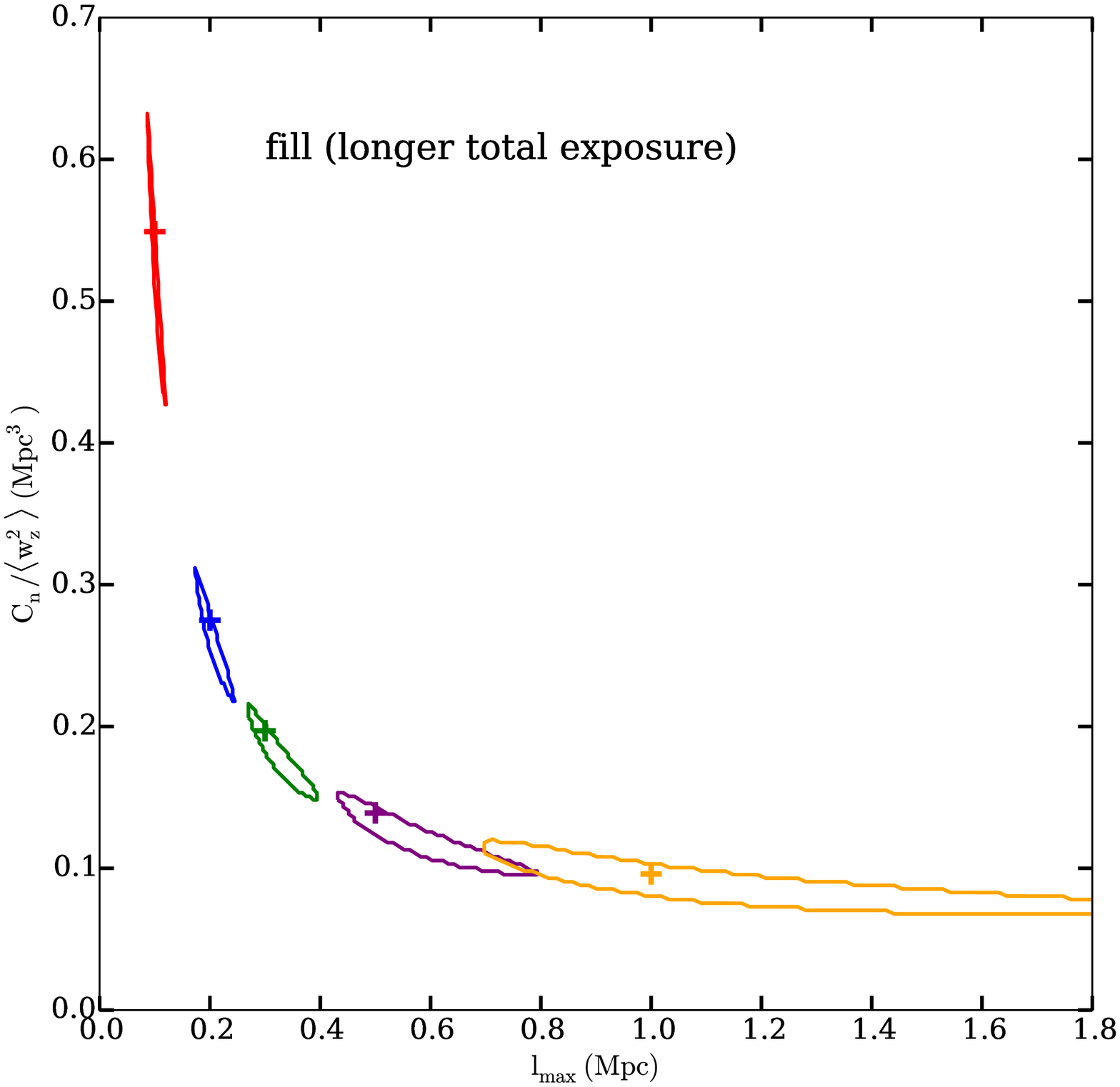}
\caption{1-parameter 68\% confidence limits on the $l_{\rm max}$ and $C_n$ parameters for four different pointing configurations and different values of $l_{\rm max}$. Crosses mark the positions of the true parameter values.}
\label{fig:lmax_constraints}
\end{center}
\end{figure*}

\subsection{Estimating the Injection Scale of Turbulence}\label{sec:estimating_inj_scale}

Our results indicate that a measurement of the 2D velocity structure function of the Coma cluster by {\it Astro-H} may be useful for determining the injection scale of the turbulent motions. To determine more precisely the constraints that future observations can put on this parameter, we jointly fit the model for the structure function (Equation \ref{eqn:ps_to_sf}) and the average velocity dispersion within the total configuration (Equation \ref{eqn:avg_vel_disp}) for the parameters $C_n$ (the power spectrum normalization) and $l_{\rm max}$ (the injection scale) to the mean structure function and average velocity dispersion computed for each pointing configuration. Note that the velocity dispersion, which we have used to define the Mach number ${\cal M}_z$, provides an additional constraint on the parameters $C_n$ and $l_{\rm max}$, but is still dependent on both (see Equation \ref{eqn:avg_vel_disp}).

We perform the fit by setting up a finely spaced 2D grid in $C_n$ and $l_{\rm max}$, and for each grid point we calculate the $\chi^2$ statistic. We compute the 1-$\sigma$ errors on the structure function and average velocity dispersion from our sample of realizations of each, which include both measurement and ``cosmic variance'' errors. All other parameter values are held fixed, at $\alpha = -11/3$ and $l_{\rm min} = 20$~kpc (other physically sensible values for these parameters will not change our conclusions, given our insensitivity to these parameters, see Sections \ref{sec:small_scale} and \ref{sec:slopes}). Under these constraints, for a measured value of the average velocity dispersion $\langle{w_z}\rangle$ ($\propto {\cal M}_z$), a given value of either $C_n$ or $l_{\rm max}$ uniquely determines the value of the other parameter.

Figure \ref{fig:lmax_constraints} shows plots of the 1-parameter 68\% ($\chi^2 < \chi^2_{\rm min}+1$) confidence regions of the parameters $l_{\rm max}$ and $C_n$, for ${\cal M}_z = 0.3$ and different values of the true $l_{\rm max}$, for the different pointing configurations. Table \ref{tab:inj_scale} shows the best-fit $l_{\rm max}$ and 1-$\sigma$ errors for each of the configurations.

\input tab2.tex

\begin{figure}
\begin{center}
\includegraphics[width=0.49\textwidth]{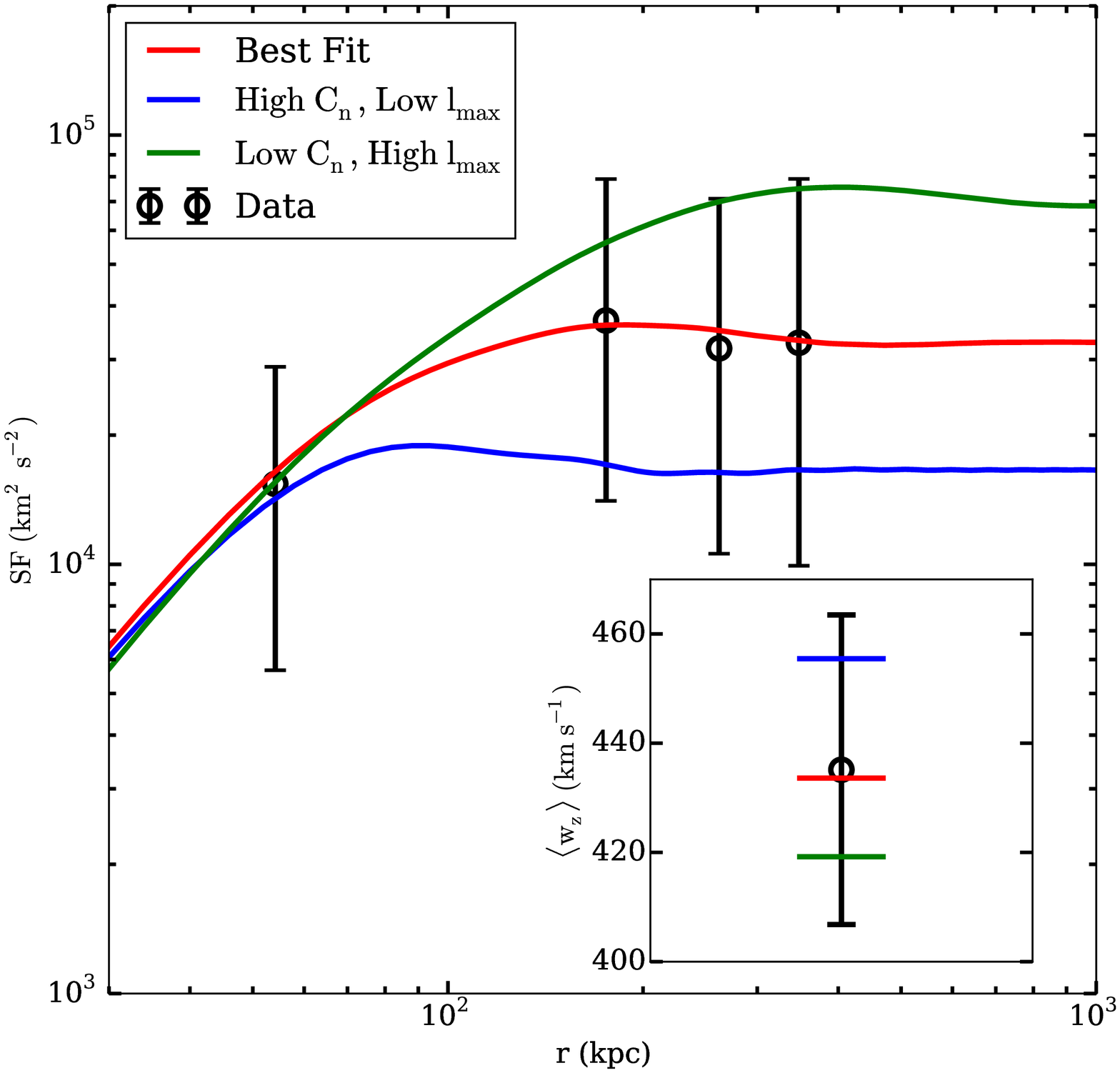}
\caption{Best-fit and example curves with extreme parameter values (taken from the extreme ends of the {\it 2-parameter} 68\% confidence regions) compared to the data for the ``big cross'' configuration, ${\cal M}_z = 0.3$, and $l_{\rm max} = 300$~kpc. The main panel shows the structure function, while the inset shows the velocity dispersion.}
\label{fig:best_fits}
\end{center}
\end{figure}

In general, we find that there is a parameter degeneracy between the normalization $C_n$ and the injection scale $l_{\rm max}$ for a given ${\cal M}_z$ (see the long banana-shaped confidence regions in Figure \ref{fig:lmax_constraints}). Figure \ref{fig:best_fits} illustrates this, by showing an example best-fit curve and two curves at extreme values of the parameters $C_n$ and $l_{\rm max}$, at the extreme ends of the two-parameter 68\% confidence region of the data ($\chi_{\rm min}^2+2.3$), both of which fit the structure function and the velocity dispersion. This figure illustrates clearly how a lower $C_n$ is compensated by a larger $l_{\rm max}$, and vice-versa.

The ``strip'' configuration (upper-left panel) does the poorest job of constraining these parameters, as shown by the overlapping confidence regions in Figure \ref{fig:lmax_constraints}. The ``big cross'' configuration (upper-right panel) fares better, due to the larger number of baselines at large distance scales, and provides tighter limits on $l_{\rm max}$. As seen in Figure \ref{fig:lmax_constraints} and Table \ref{tab:inj_scale}, constraining $l_{\rm max}$ at values larger than those that are covered by our region of interest ($\sim$400~kpc wide) is difficult. Such cluster-scale eddies might be caused by a recent major merger.

It is readily apparent from Figure \ref{fig:lmax_constraints} (and not unexpected) that the ``fill'' configuration with the same exposure time per pointing (and much longer total exposure, lower-right panel) does the best job of constraining the injection scale and normalization of the turbulent cascade. However, this is a very expensive survey. The ``checkerboard'' configuration (lower-left panel), with its lower exposure time per pointing but more pointings, offers a modest improvement over ``big cross'' in terms of constraining the injection scale. However, it relies on measuring velocities based on a small number of line counts, the possibility of which needs to be demonstrated with real data.

\subsection{Multiple Injection Scales}\label{sec:mult_scales}

In reality, the distribution of turbulent velocities in the Coma cluster is likely to be more complicated than the simple power-law models we have investigated in this work. One likely possibility, motivated by cosmological simulations of merging clusters, is that there will be more than one scale at which significant driving of turublent motions is occurring. In the Coma cluster, there is an ongoing cluster merger which is driving turbulence at large scales (a few hundred kpc), whereas the motion of galaxies within the ICM will stir the gas at scales of 100 kpc or less.

Using numerical simulations, \citet{yoo14} explored the effect of driving turbulence on multiple length scales. They found in the case of driving on two scales that the relative height of the peaks of the power spectra and the separation of the driving scales is crucial to whether or not the two components can be distinguished. In particular, it will be easiest to discern the presence of driving on two length scales if the two peaks are roughly equal in height and the driving scales are well-separated.

\begin{figure*}
\begin{center}
\includegraphics[width=0.97\textwidth]{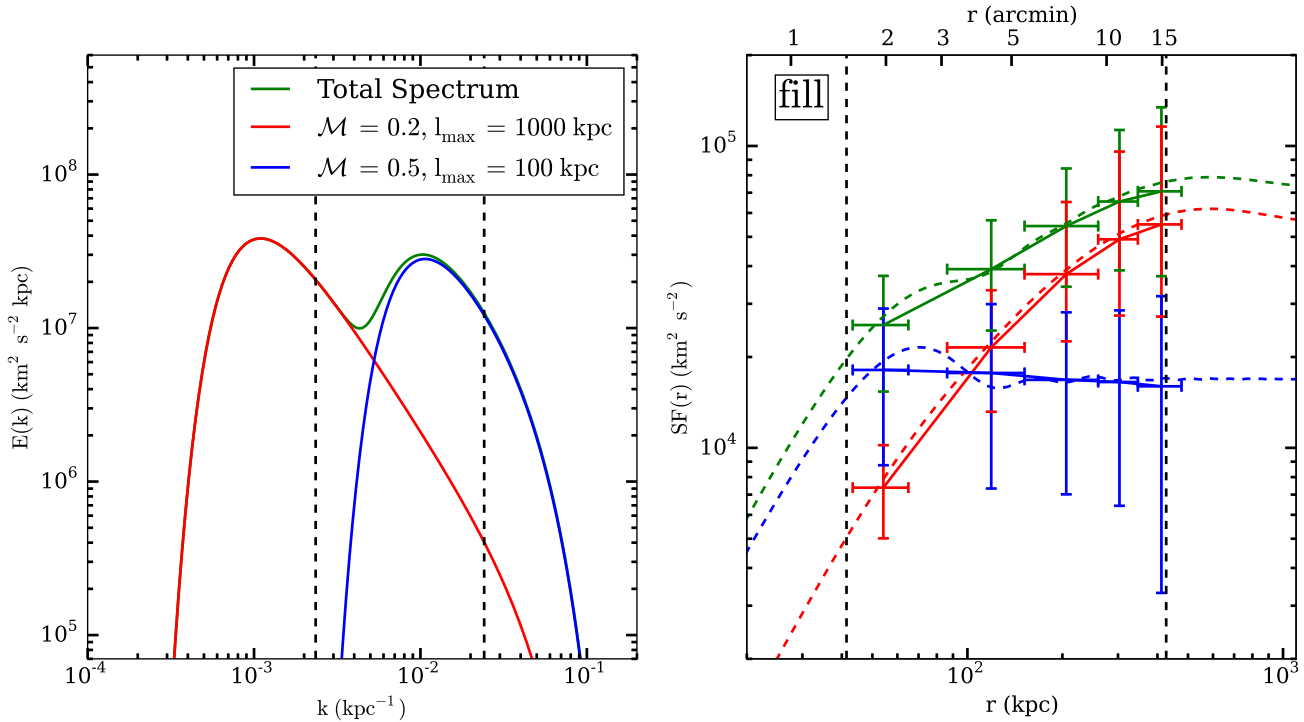}
\caption{Driving turbulence on more than one scale. Left panel: Energy spectra for a model where the total spectrum is a sum of two components with different driving scales. Right panel: Expected and computed structure function for the two-scale model. Red and blue curves show the individual model components, whereas the green curve shows the total model.}
\label{fig:two_scales}
\end{center}
\end{figure*}

To investigate the ability of {\it Astro-H} to distinguish multiple scales, we modeled an input power spectrum that is the sum of two single-injection scale models (shown in the left panel of Figure \ref{fig:two_scales}), one with ${\cal M}$ = 0.5 and an injection scale of $l_{\rm max}$ = 100~kpc and another with ${\cal M}$ = 0.2 and an injection scale of $l_{\rm max}$ = 1000~kpc. The green curve shows the total model, and the red and blue curves show the individual subcomponents of this model. What results is an energy spectrum with two resolvable peaks of approximately the same height, separated enough in length scale so that they are distinct. We use this input power spectrum to generate velocities in the same manner as our previously described models.

The right panel of Figure \ref{fig:two_scales} shows the corresponding structure function for this model. The dashed green line shows the expected curve. The structure function has a flattening at intermediate $r$ that corresponds to the smaller injection scale, then an increase followed by another flattening corresponding to the larger injection scale. The solid lines with error bars show the computed structure functions from the ``fill'' pointing configuration, chosen to maximize the probability that the two injection scales can be discerned. However, even in this case, due to the finite resolution the computed structure function is unable to resolve the transition region between the two subcomponents at an angular scale of approximately 3.5' (length scale of 100~kpc) between the two injection scales, and instead the behavior of the combined model appears indistinguishable from a power-law within scales 40~kpc~$\simlt r \simlt 300$~kpc. The slope appears shallower than those in the single-injection models, since the reduction in the structure function with decreasing $r$ from the $l_{\rm max} = 1000$~kpc component is partially offset by the contribution from the $l_{\rm max} = 100$~kpc component.

This demonstrates that distinguishing between multiple injection scales will be difficult, not only due to the cosmic variance and measurement uncertainty, but also from the limited range of length scales that {\it Astro-H} will be able to sample. It is also clear from Figure \ref{fig:two_scales} that a mission with a higher spatial resolution, such as {\it Athena} or {\it Smart-X}, would present a better opportunity for discerning the two driving scales, since the slope of the lower-injection scale model steepens at angular scales less than 1.5' (length scales $\simlt 40$~kpc).

\section{Summary}\label{sec:conc}

We have performed simulations of turbulent velocity fields in the Coma cluster, and constructed the 2D structure function of the line of sight velocity field (line shift). These simulations bear directly on future observations of the Coma cluster by {\it Astro-H}. We have simulated a number of realistic pointing configurations and models for the underlying velocity field. Our main results are as follows:

\begin{itemize}
\item It will not be feasible to distinguish between structure functions with different dissipation scales, regardless of the number and layout of the pointings used, unless this scale is implausibly large (on the order of 100s of kpc).
\item It will be possible to measure the turbulence injection scale, for any of the mapping configurations examined in this work, though some configurations resolve some length scales better than others. However, this is likely only feasible for turbulence where the Mach number of the line of sight velocity dispersion is ${\cal M}_z \simgt 0.2$, unless the systematic uncertainties are much better than anticipated.
\item It will not be feasible to distinguish between different power-law slopes in the inertial range of the power spectrum, unless the slopes are significantly steeper or shallower than suggested by physically motivated models of turbulence.
\item Since the structure functions for different injection scales are best-distinguished by their behavior at large distance scales, the ``big cross'' pointing configuration is preferable to the ``strip'' pointing configuration, as it samples larger length scales better at the expense of intermediate scales. However, we find it is advantageous to split the same total exposure among a larger number of offsets. The ``checkerboard'' configuration places tighter constraints on the injection scale, though these improvements are modest. We therefore recommend that Coma should be mapped with either the ``big cross'' pointing at $\sim$100~ks per pointing, or with the additional four pointings required to make the ``checkerboard'' configuration at $\sim$55~ks per pointing. The latter configuration will reduce the cosmic variance error on the structure function, but it requires confidence (beyond the formal statistical sense) in velocity measurements derived from such a low number of counts.
\end{itemize}

Other statistics constructed from the measured velocity field of Coma may also be of use in determining the properties of the velocity field. Z12 suggested using the ratio of the root mean square of the line shift to the velocity dispersion as a function of length scale to constrain the injection scale of turbulence, a measure similar to our joint fit of the structure function and the average velocity dispersion in Section \ref{sec:estimating_inj_scale}. We have also not examined the spatial dependence of the velocity dispersion (line broadening), which may place additonal constraints on the turbulent power spectrum.

Furthermore, the non-Gaussian shape of the turbulently broadened line profiles also encodes information on the turbulent power spectrum \citep[][]{ino03}, but to access this information, much longer exposures will be required. For well-resolved clusters such as Coma, we can constrain the power spectrum with cheaper, spatially resolved measurements, as shown in this paper. However, for those clusters where the {\it Astro-H} angular resolution will be in, the information in the shape of the line profile may be essential.

Finally, though similar analyses may be performed with other clusters, it will be more difficult. We have chosen Coma because its large core radius and lack of a central AGN make it a relatively clean setup for this kind of analysis. Other nearby systems, such as Virgo and Perseus, have the requisite spatial resolution, but have steep entropy gradients and a large dynamic range in density and pressure within the core region. Additionally, they are both sources of AGN activity and possess sloshing cold fronts, which will limit the locations at which we can expect to get a measurement of the velocity field that is not contaminated by these effects. Perhaps most importantly, the simplifying assumptions of homogeneous and isotropic turbulence made in this work will have to be relaxed, since the significant stratifcation of the cluster atmosphere in these systems will prevent large eddies from developing in the radial direction, while permitting them to develop in the tangential direction \citep{zhu14a,zhu15}. Z12 showed that the velocity dispersion as a function of projected radius from the cluster center is effectively a measurement of the structure function, though we must add a caveat that for clusters with highly stratified atmospheres the differences in scales probed by this method are significantly affected by differences in the power spectrum itself as a function of radius.

Our results have shown that {\it Astro-H} will be able, for the first time, to directly measure the Mach number and the injection scale of the turbulent velocity field of a galaxy cluster, in this case the nearby Coma cluster. These are important quantities, as they (a) determine the diffusion rate for everything advected by the gas, including cosmic rays, metals, etc., and (b) the rate of energy flow down the turbulent cascade, which ends up dissipated into heat and nonthermal components. Resolving other aspects of ICM turbulence that have bearing on the underlying physics will require a mission with finer angular resolution and larger effective area, such as the upcoming {\it Athena}\footnote{\url{http://www.the-athena-x-ray-observatory.eu/}} or the proposed {\it Smart-X}\footnote{\url{http://smart-x.cfa.harvard.edu/}}. These include resolving the turbulent dissipation scale and constraining the spectral index of the turbulent power spectrum, both of which depend on the microphysics of the plasma and the driving mechanism of the gas motions. The lessons learned in this work can be applied to future investigations assuming the characteristics of any future X-ray instrument.

\acknowledgments
JAZ thanks Mark Bautz, Eric Miller, Fred Baganoff, Catherine Grant, and Michael McDonald for several useful discussions. This research made use of AstroPy, a community-developed core Python package for Astronomy \citep{apy13}. JAZ acknowledges support from NASA though subcontract SV2-8203 to MIT from the Smithsonian Astrophysical Observatory, as well as the Astrophysics Theory Program Award Number 12-ATP12-0159.

{}

\appendix

\section{A. Calculation of the Normalization of the 2D Power Spectrum}\label{sec:norm}

Section \ref{sec:2d_ps_and_sf} indicated the relationship of the velocity power spectrum in two and three dimensions can be approximated by a normalization constant $K$, given by
\begin{equation}
K = \displaystyle\int{P_\epsilon(k_z;x,y)dk_z}
\end{equation}
where $P_\epsilon(x,y;k_z)$ is the power spectrum of the normalized emission $\omega({\bf x})$ for a given position $(x,y)$ on the sky. Our choice of $\beta = 2/3$ in Equation \ref{eqn:beta_model} simplifies the required integrations considerably. Setting $R^2 = x^2+y^2$ and $c^2 = r_c^2 + R^2$, we can rewrite Equation \ref{eqn:EM} as
\begin{equation}
\omega({\bf x}) = \frac{\left[1+\left(\frac{r}{r_c}\right)^2\right]^{-3\beta}}{\displaystyle\int_{-\infty}^\infty\left[1+\left(\frac{r}{r_c}\right)^2\right]^{-3\beta}dz} = \frac{\frac{1}{(c^2+z^2)^2}}{\displaystyle\int_{-\infty}^\infty\frac{dz}{(c^2+z^2)^2}} = \frac{2c^3}{\pi}\frac{1}{(c^2+z^2)^2}
\end{equation}
the Fourier transform of $\omega$ is therefore given by
\begin{equation}
\tilde{\omega}(k_z;x,y) = \frac{2c^3}{\pi}\displaystyle\int_{-\infty}^\infty\frac{1}{(c^2+z^2)^2}e^{-2\pi{ik_z}z}dz = e^{-2\pi{c|k_z|}}(2\pi{c|k_z|}+1)
\end{equation}
since $\omega$ is even the imaginary part of the Fourier transform is zero. Finally, setting $P_\epsilon(x,y;k_z) = |\omega|^2$, we have:
\begin{equation}
K = \displaystyle\int_{-\infty}^\infty{P_\epsilon(k_z;x,y)dk_z} = \displaystyle\int_{-\infty}^\infty{e^{-4\pi{c|k_z|}}(2\pi{c|k_z|}+1)^2}dk_z = \frac{5}{4\pi\sqrt{R^2+r_c^2}}
\end{equation}
The average value of $K$ over our region of interest (a radius of approximately 200~kpc) is $\sim$1~Mpc$^{-1}$.

\section{B. Derivation of the Structure Function}\label{sec:sf_derivation}

The structure function $SF(r)$ (Equation \ref{eqn:2d_sf}) can be rewritten as:
\begin{equation}\label{eqn:2d_sf_expanded}
SF(r) = \langle{\bar{v}_z^2(\boldsymbol{\chi}+{\bf r})}\rangle + \langle{\bar{v}_z^2(\boldsymbol{\chi})}\rangle - 2\langle{\bar{v}_z(\boldsymbol{\chi}+{\bf r})\bar{v}_z(\boldsymbol{\chi})}\rangle.
\end{equation}

Via the Weiner-Khinchin theorem, the inverse Fourier transform of the 2D power spectrum $P_{\rm 2D}(\bf{k})$ is the velocity autocorrelation:
\begin{equation}
\Gamma(r) = \langle{\bar{v}_z(\boldsymbol{\chi}+{\bf r})\bar{v}_z(\boldsymbol{\chi})}\rangle = \int{e^{{2\pi}i\bf{k}\cdot{\bf r}}}{P_{\rm 2D}}({\bf k})d^2{\bf k}
\end{equation}
Since the velocity field is stationary, the first two terms of Equation \ref{eqn:2d_sf_expanded} are:
\begin{equation}
\langle{\bar{v}_z^2(\boldsymbol{\chi}+{\bf r})}\rangle = \langle{\bar{v}_z^2(\boldsymbol{\chi})}\rangle = \Gamma(0) = \int{P_{\rm 2D}}({\bf k})d^2{\bf k},
\end{equation}
yielding
\begin{equation}
SF(r) = 2(\Gamma(0)-\Gamma(r)) = 2\int\left[1-e^{{2\pi}i{\bf k}\cdot{\bf r}}\right]P_{\rm 2D}({\bf k})d^2{\bf k} = 2\int\left[1-\cos(2\pi{\bf k}\cdot{\bf r})\right]P_{\rm 2D}({\bf k})d^2{\bf k}.
\end{equation}
where the last equality follows since the integral is even. If we take the integration in polar coordinates and use Parseval's integral\citep{arf05}, we finally have
\begin{equation}
SF(r) = {4\pi}\int_0^\infty{\left[1-J_0({2\pi}kr)\right]P_{\rm 2D}(k)kdk}
\end{equation}
where $J_0$ is a Bessel function of the first kind of order zero.

\section{C. Calculation of the Bias on the Structure Function Due to Measurement Errors}\label{sec:bias}

For a single pair of velocities $v_i$, $v_j$ separated by a distance $r_{ij}$, we define
\begin{equation}
s_{ij} = (v_i-v_j)^2
\end{equation}
The structure function $SF(r)$ is an average over the $s_{ij}$ at a given length scale $r$. The observed velocities will include measurement errors, which can be modeled as additions to the pointwise velocities, giving the observed structure function value as
\begin{equation}
s'_{ij} = [(v_i+\delta{v_i})-(v_j+\delta{v_j})]^2
\end{equation}
where the $\delta{v_i}$, $\delta{v_j}$ are normally distributed with zero mean and variances $\sigma_i^2$, $\sigma_j^2$. Expanding and rearranging terms, we have
\begin{equation}
s'_{ij} = (v_i-v_j)^2 + \delta{v_i}^2 + \delta{v_j}^2 - 2\delta{v_i}\delta{v_j} + 2\delta{v_i}(v_i-v_j) - 2\delta{v_j}(v_i-v_j)
\end{equation}
The total observed structure function is an average over the $s'_{ij}$ for a given length scale $r$, yielding
\begin{eqnarray}
\langle{s'_{ij}}\rangle_r &=& \langle(v_i-v_j)^2\rangle_r + \langle\delta{v_i}^2\rangle_r +
\langle\delta{v_j}^2\rangle_r - \langle2\delta{v_i}\delta{v_j}\rangle_r + \langle2\delta{v_i}(v_i-v_j)\rangle_r - \langle2\delta{v_j}(v_i-v_j)\rangle_r \\
\nonumber &=& \langle(v_i-v_j)^2\rangle_r + \langle\delta{v_i}^2\rangle_r +
\langle\delta{v_j}^2\rangle_r - 2\langle\delta{v_i}\rangle_r\langle\delta{v_j}\rangle_r + 2\langle\delta{v_i}\rangle_r\langle(v_i-v_j)\rangle_r - 2\langle\delta{v_j}\rangle_r\langle(v_i-v_j)\rangle_r
\end{eqnarray}
where the second line follows from the statistical independence of the measurement errors from different locations and from the value of the velocity. Since $\langle\delta{v_i}\rangle_r$ = $\langle\delta{v_j}\rangle_r$ = 0, this reduces to
\begin{eqnarray}
\langle{s'_{ij}}\rangle_r &=& \langle(v_i-v_j)^2\rangle_r + \langle\delta{v_i}^2\rangle_r +
\langle\delta{v_j}^2\rangle_r \\
\nonumber &=& \langle{s_{ij}}\rangle_r + \sigma_i^2 + \sigma_j^2
\end{eqnarray}
Finally, since $\sigma_i$ = $\sigma_j$, and assuming at a given point $i$ $\sigma_i^2 = \sigma_{\rm stat}^2 + \sigma_{\rm sys}^2$, the sum in quadrature of the statistical and systematic errors, we have
\begin{equation}
\langle{s'_{ij}}\rangle_r = \langle{s_{ij}}\rangle_r + 2\sigma_{\rm stat}^2 + 2\sigma_{\rm sys}^2
\end{equation}
For this work, we have assumed that velocity differences on the smallest length scales (less than $\approx$~1.5') have zero systematic error, to simplify the calculation of the bias. Though this is only strictly true for velocity differences on the same pointing, we find that including systematic errors on velocity differences at this small scale has little effect on our results, due to the fact that the number of differences within pointings is much larger than the number of differences across pointings for any of our setups.

\end{document}

%% file: tab1.tex
\begin{table*}
\tabletypesize{\scriptsize}
\caption{Statistical Errors on the Line Shift and Width\label{tab:stat_err}}
\begin{center}
\begin{tabular}{cccccc}
\hline
\hline
${\cal M}_z$ & $w_{z,0}$ (km/s) & $\sigma_{\rm stat,shift}$ (km/s) & $\sigma_{\rm stat,shift}/w_{z,0}$ (\%) & $\sigma_{\rm stat,width}$ (km/s) & $\sigma_{\rm stat,width}/w_{z,0}$ (\%) \\
\hline
0.1 & 146 & 33 & 22.6 & 50 & 34.2 \\
0.2 & 292 & 59 & 20.2 & 71 & 24.3 \\
0.3 & 438 & 86 & 19.6 & 97 & 19.6 \\
0.5 & 729 & 142 & 19.5 & 153 & 21.0 \\
\hline
\end{tabular}
\end{center}
\end{table*}

%% file: tab2.tex
\renewcommand{\arraystretch}{1.5}
\setlength{\tabcolsep}{0.4em}
\begin{table}[thp]
\tabletypesize{\scriptsize}
\caption{Constraints on $l_{\rm max}$\label{tab:inj_scale}}
\begin{center}
\begin{tabular}{lllll}
\hline
\hline
$l_{\rm max}$ (kpc) & \multicolumn{4}{c}{Estimated $l_{\rm max}$ (kpc)} \\
 & strip & big cross & fill & checkerboard \\
\hline
100 & 105$\substack{+54 \\ -32}$ & 103$\substack{+56 \\ -36}$ & 103$\substack{+18 \\ -16}$ & 103$\substack{+46 \\ -34}$ \\
200 & 209$\substack{+133 \\ -64}$ & 205$\substack{+113 \\ -68}$ & 205$\substack{+40 \\ -32}$ & 201$\substack{+80 \\ -52}$ \\
300 & 329$\substack{+314 \\ -111}$ & 329$\substack{+203 \\ -111}$ & 322$\substack{+72 \\ -52}$ & 322$\substack{+144 \\ -85}$ \\
500 & 623$\substack{+1196 \\ -281}$ & 582$\substack{+613 \\ -211}$ & 552$\substack{+241 \\ -121}$ & 562$\substack{+442 \\ -171}$ \\
1000 & 1339$\substack{+1370 \\ -845}$ & 1208$\substack{+1239 \\ -612}$ & 1062$\substack{+889 \\ -364}$ & 1120$\substack{+1180 \\ -510}$ \\
\hline
\end{tabular}
\end{center}
\end{table}

%% file: ms.bbl
\begin{thebibliography}{}
\bibitem[Applegate et al.(2014)]{app14} Applegate, D.~E., von der Linden, A., Kelly, P.~L., et al.\ 2014, \mnras, 439, 48
\bibitem[Arfken \& Weber(2005)]{arf05} Arfken, G.~B., \& Weber, H.~J.\ 2005, ``Mathematical Methods for Physicists'' 6th Ed. Boston: Elsevier, 2005.
\bibitem[Astropy Collaboration et al.(2013)]{apy13} Astropy Collaboration, Robitaille, T.~P., Tollerud, E.~J., et al.\ 2013, \aap, 558, A33
\bibitem[Banerjee \& Sharma(2014)]{ban14} Banerjee, N., \& Sharma, P.\ 2014, \mnras, 443, 687
\bibitem[Braginskii(1965)]{bra65} Braginskii, S.~I.\ 1965, Reviews of Plasma Physics, 1, 205
\bibitem[Brandenburg \& Lazarian(2013)]{bra13} Brandenburg, A., \& Lazarian, A.\ 2013, \ssr, 178, 163
\bibitem[Briel et al.(1992)]{bri92} Briel, U.~G., Henry, J.~P., \& Boehringer, H.\ 1992, \aap, 259, L31
\bibitem[Brown \& Rudnick(2011)]{bro11} Brown, S., \& Rudnick, L.\ 2011, \mnras, 412, 2
\bibitem[Brunetti \& Lazarian(2007)]{bru07} Brunetti, G., \& Lazarian, A.\ 2007, \mnras, 378, 245
\bibitem[Cavaliere \& Fusco-Femiano(1976)]{cav76} Cavaliere, A., \& Fusco-Femiano, R.\ 1976, \aap, 49, 137
\bibitem[Cavaliere \& Fusco-Femiano(1978)]{cav78} Cavaliere, A., \& Fusco-Femiano, R.\ 1978, \aap, 70, 677
\bibitem[Churazov et al.(2004)]{chu04} Churazov, E., Forman, W., Jones, C., Sunyaev, R., B\"{o}hringer, H.\ 2004, \mnras, 347, 29
\bibitem[Churazov et al.(2012)]{chu12} Churazov, E., Vikhlinin, A., Zhuravleva, I., et al.\ 2012, \mnras, 421, 1123
\bibitem[de Plaa et al.(2012)]{dep12} de Plaa, J., Zhuravleva, I., Werner, N., et al.\ 2012, \aap, 539, A34
\bibitem[Dennis \& Chandran(2005)]{den05} Dennis, T.~J., \& Chandran, B.~D.~G.\ 2005, \apj, 622, 205
\bibitem[Donnert et al.(2013)]{don13} Donnert, J., Dolag, K., Brunetti, G., \& Cassano, R.\ 2013, \mnras, 429, 3564
\bibitem[En{\ss}lin et al.(2011)]{ens11} En{\ss}lin, T., Pfrommer, C., Miniati, F., \& Subramanian, K.\ 2011, \aap, 527, A99
\bibitem[Evrard et al.(1996)]{evr96} Evrard, A.~E., Metzler, C.~A., \& Navarro, J.~F.\ 1996, \apj, 469, 494
\bibitem[Fabian et al.(2003)]{fab03} Fabian, A.~C., Sanders, J.~S., Crawford, C.~S., et al.\ 2003, \mnras, 344, L48
\bibitem[Fujita et al.(2004)]{fuj04} Fujita, Y., Matsumoto, T., \& Wada, K.\ 2004, \apjl, 612, L9
\bibitem[Hughes(1989)]{hug89} Hughes, J.~P.\ 1989, \apj, 337, 21
\bibitem[Inogamov \& Sunyaev(2003)]{ino03} Inogamov, N.~A., \& Sunyaev, R.~A.\ 2003, Astronomy Letters, 29, 791
\bibitem[Kolmogorov(1941)]{kol41} Kolmogorov, A.\ 1941, Akademiia Nauk SSSR Doklady, 30, 301
\bibitem[Kunz et al.(2014)]{kun14} Kunz, M.~W., Schekochihin, A.~A., \& Stone, J.~M.\ 2014, PhRvL, 112, 205003
\bibitem[Lazarian(2006a)]{laz06a} Lazarian, A.\ 2006, Astronomische Nachrichten, 327, 609
\bibitem[Lazarian(2006b)]{laz06b} Lazarian, A.\ 2006, \apjl, 645, L25
\bibitem[Lazarian \& Beresnyak(2006)]{laz06} Lazarian, A., \& Beresnyak, A.\ 2006, \mnras, 373, 1195
\bibitem[Mahdavi et al.(2013)]{mah13} Mahdavi, A., Hoekstra, H., Babul, A., et al.\ 2013, \apj, 767, 116
\bibitem[Markevitch \& Vikhlinin(2007)]{MV07} Markevitch, M., \& Vikhlinin, A.\ 2007, \physrep, 443, 1
\bibitem[Miniati\& Beresnyak(2015)]{min15} Miniati, F., \& Beresnyak, A.\ 2015, \nat, 523, 59
\bibitem[Mogavero \& Schekochihin(2014)]{mog14} Mogavero, F., \& Schekochihin, A.~A.\ 2014, \mnras, 440, 3226
\bibitem[Murgia et al.(2004)]{mur04} Murgia, M., Govoni, F., Feretti, L., et al.\ 2004, \aap, 424, 429
\bibitem[Nagai et al.(2007)]{nag07} Nagai, D., Vikhlinin, A., \& Kravtsov, A.~V.\ 2007, \apj, 655, 98
\bibitem[Nelson et al.(2014)]{nel14} Nelson, K., Lau, E.~T., \& Nagai, D.\ 2014, \apj, 792, 25
\bibitem[Neumann et al.(2003)]{neu03} Neumann, D.~M., Lumb, D.~H., Pratt, G.~W., \& Briel, U.~G.\ 2003, \aap, 400, 811
\bibitem[Nulsen \& McNamara(2013)]{nul13} Nulsen, P.~E.~J., \& McNamara, B.~R.\ 2013, Astronomische Nachrichten, 334, 386
\bibitem[Piffaretti \& Valdarnini(2008)]{pif08} Piffaretti, R., \& Valdarnini, R.\ 2008, \aap, 491, 71
\bibitem[Pinto et al.(2015)]{pin15} Pinto, C., Sanders, J.~S., Werner, N., et al.\ 2015, \aap, 575, A38
\bibitem[Poole et al.(2006)]{poo06} Poole, G.~B., Fardal, M.~A., Babul, A., McCarthy, I.~G., Quinn, T., \& Wadsley, J.\ 2006, \mnras, 373, 881
\bibitem[Rasia et al.(2006)]{ras06} Rasia, E., Ettori, S., Moscardini, L., et al.\ 2006, \mnras, 369, 2013
\bibitem[Rebusco et al.(2006)]{reb06} Rebusco, P., Churazov, E., B{\"o}hringer, H., \& Forman, W.\ 2006, \mnras, 372, 1840
\bibitem[Ricker \& Sarazin(2001)]{ric01} Ricker, P.~M., \& Sarazin, C.~L.\ 2001, \apj, 561, 621
\bibitem[Roediger et al.(2013)]{rod13} Roediger, E., Kraft, R.~P., Forman, W.~R., et al.\ 2013, \apj, 764, 60
\bibitem[Ruszkowski et al.(2007)]{rus07} Ruszkowski, M., En{\ss}lin, T.~A., Br{\"u}ggen, M., Heinz, S., \& Pfrommer, C.\ 2007, \mnras, 378, 662
\bibitem[Ruszkowski \& Oh(2010)]{rus10} Ruszkowski, M., \& Oh, S.~P.\ 2010, \apj, 713, 1332
\bibitem[Sanders et al.(2011)]{san11} Sanders, J.~S., Fabian, A.~C., \& Smith, R.~K.\ 2011, \mnras, 410, 1797
\bibitem[Sanders \& Fabian(2013)]{san13a} Sanders, J.~S., \& Fabian, A.~C.\ 2013, \mnras, 429, 2727
\bibitem[Sanders et al.(2013)]{san13b} Sanders, J.~S., Fabian, A.~C., Churazov, E., et al.\ 2013, Science, 341, 1365
\bibitem[Schekochihin \& Cowley(2006)]{sch06} Schekochihin, A.~A., \& Cowley, S.~C.\ 2006, Physics of Plasmas, 13, 056501
\bibitem[Schuecker et al.(2004)]{sch04} Schuecker, P., Finoguenov, A., Miniati, F., B{\"o}hringer, H., \& Briel, U.~G.\ 2004, \aap, 426, 387
\bibitem[Spitzer(1962)]{spi62} Spitzer, L.\ 1962, Physics of Fully Ionized Gases, New York: Interscience (2nd edition), 1962,
\bibitem[Takahashi et al.(2012)]{tak12} Takahashi, T., Mitsuda, K., Kelley, R., et al.\ 2012, \procspie, 8443, 84431Z
\bibitem[Vazza et al.(2009)]{vaz09} Vazza, F., Brunetti, G., Kritsuk, A., et al.\ 2009, \aap, 504, 33
\bibitem[Vazza et al.(2010)]{vaz10} Vazza, F., Gheller, C., \& Brunetti, G.\ 2010, \aap, 513, A32
\bibitem[von der Linden et al.(2014)]{vdl14} von der Linden, A., Mantz, A., Allen, S.~W., et al.\ 2014, \mnras, 443, 1973
\bibitem[Werner et al.(2009)]{wer09} Werner, N., Zhuravleva, I., Churazov, E., et al.\ 2009, \mnras, 398, 23
\bibitem[Yoo \& Cho(2014)]{yoo14} Yoo, H., \& Cho, J.\ 2014, \apj, 780, 99
\bibitem[Zhang et al.(2010)]{zha10} Zhang, Y.-Y., Okabe, N., Finoguenov, A., et al.\ 2010, \apj, 711, 1033
\bibitem[Zhuravleva et al.(2012)]{zhu12} Zhuravleva, I., Churazov, E., Kravtsov, A., \& Sunyaev, R.\ 2012, \mnras, 422, 2712
\bibitem[Zhuravleva et al.(2013)]{zhu13} Zhuravleva, I., Churazov, E., Sunyaev, R., et al.\ 2013, \mnras, 435, 3111
\bibitem[Zhuravleva et al.(2014a)]{zhu14a} Zhuravleva, I., Churazov, E.~M., Schekochihin, A.~A., et al.\ 2014, \apjl, 788, L13
\bibitem[Zhuravleva et al.(2014b)]{zhu14} Zhuravleva, I., Churazov, E., Schekochihin, A.~A., et al.\ 2014, \nat, 515, 85
\bibitem[Zhuravleva et al.(2015)]{zhu15} Zhuravleva, I., Churazov, E., Ar{\'e}valo, P., et al.\ 2015, \mnras, 450, 4184
\bibitem[ZuHone et al.(2010)]{zuh10} ZuHone, J.~A., Markevitch, M., \& Johnson, R.~E.\ 2010, \apj, 717, 908 (ZMJ10)
\bibitem[ZuHone et al.(2011)]{zuh11} ZuHone, J.~A., Markevitch, M., \& Lee, D.\ 2011, \apj, 743, 16
\bibitem[ZuHone et al.(2013)]{zuh13} ZuHone, J.~A., Markevitch, M., Brunetti, G., \& Giacintucci, S.\ 2013, \apj, 762, 78
\bibitem[ZuHone et al.(2015)]{zuh15} ZuHone, J.~A., Kunz, M.~W., Markevitch, M., Stone, J.~M., \& Biffi, V.\ 2015, \apj, 798, 90
\end{thebibliography}
